%% file: main.tex
\documentclass[sigconf]{acmart}

\AtBeginDocument{%
  \providecommand\BibTeX{{%
    \normalfont B\kern-0.5em{\scshape i\kern-0.25em b}\kern-0.8em\TeX}}}

\copyrightyear{2021}
\acmYear{2021}
\setcopyright{acmlicensed}\acmConference[CHI '21]{CHI Conference on Human Factors in Computing Systems}{May 8--13, 2021}{Yokohama, Japan}
\acmBooktitle{CHI Conference on Human Factors in Computing Systems (CHI '21), May 8--13, 2021, Yokohama, Japan}
\acmPrice{15.00}
\acmDOI{10.1145/3411764.3445695}
\acmISBN{978-1-4503-8096-6/21/05}

\acmConference[CHI '21]{CHI '21: ACM CHI Conference on Human Factors in Computing Systems}{May 08--13, 2021}{Yokohama, Japan}
\acmBooktitle{CHI '21:  ACM CHI Conference on Human Factors in Computing Systems, May 08--13, 2021, Yokohama, Japan}
\acmPrice{15.00}
\acmISBN{978-1-4503-XXXX-X/18/06}

\usepackage{balance}       
\usepackage{graphics}      
\usepackage[T1]{fontenc}   
\usepackage{mathptmx}
\usepackage{color}
\usepackage{booktabs}
\usepackage{textcomp}
\usepackage{multicol}
\usepackage{mwe}

\usepackage{microtype}        
\usepackage{ccicons}          

\usepackage{todonotes}

\usepackage{enumitem}
\setitemize{noitemsep,topsep=0pt,parsep=0pt,partopsep=0pt}

\def\plaintitle{Not Now, Ask Later: Users Weaken Their Behavior Change Regimen Over Time, But Expect To Re-Strengthen It Imminently}

\def\emptyauthor{}
\def\plainkeywords{Behavior change; distractions and interruptions.}

\title[Not Now, Ask Later]{Not Now, Ask Later: Users Weaken Their Behavior Change Regimen Over Time, But Expect To Re-Strengthen It Imminently}

\definecolor{linkColor}{RGB}{6,125,233}
\hypersetup{%
  pdftitle={\plaintitle},
  pdfauthor={\emptyauthor},
  pdfkeywords={\plainkeywords},
  pdfdisplaydoctitle=true, 
  bookmarksnumbered,
  pdfstartview={FitH},
  colorlinks,
  citecolor=black,
  filecolor=black,
  linkcolor=black,
  urlcolor=linkColor,
  breaklinks=true,
  hypertexnames=false
}

\begin{document}


\author{Geza Kovacs}
\affiliation{
  \institution{Lilt, Inc}
  \streetaddress{Anonymous}
  \city{San Francisco}
  \state{CA}
  \country{USA}}

\author{Zhengxuan Wu}
\affiliation{
  \institution{Stanford University}
  \streetaddress{Anonymous}
  \city{Stanford}
  \state{CA}
  \country{USA}}

\author{Michael S. Bernstein}
\affiliation{
  \institution{Stanford University}
  \streetaddress{Anonymous}
  \city{Stanford}
  \state{CA}
  \country{USA}}

\begin{abstract}
How effectively do we adhere to nudges and interventions that help us control our online browsing habits?
If we have a temporary lapse and disable the behavior change system, do we later resume our adherence, or has the dam broken?
In this paper, we investigate these questions through log analyses of 8,000+ users on HabitLab, a behavior change platform that helps users reduce their time online. We find that, while users typically begin with high-challenge interventions, over time they allow themselves to slip into easier and easier interventions. Despite this, many still expect to return to the harder interventions imminently: they repeatedly choose to be asked to change difficulty again on the next visit, declining to have the system save their preference for easy interventions.
\end{abstract}

%
%



\maketitle




\input{introduction}

\input{relatedwork}

\input{researchquestions}

\input{system}

\input{study1}

\input{study2}


\input{discussion}

\input{conclusion}

\balance{}

\bibliographystyle{ACM-Reference-Format}
\bibliography{bibliography}

\end{document}

%% file: introduction.tex

\section{Introduction}

More people are working with computers~\cite{pena2002nation} and online~\cite{pabilonia2020telework, czaja2020setting, lister2019telework} than ever before, making distractions an ever-present problem~\cite{dabbish2011keep, tseng2019overcoming, mark2015focused}. While tied to well-being outcomes~\cite{burke2016relationship}, social media and other platforms also often lead to self-interruptions that people wish they are better able to control~\cite{mark2018effects,kovacs2018rotating}. Many productivity tools have emerged to combat online distractions~\cite{lyngs2019self,kovacs2019conservation}, yet keeping users adhering to interventions remains a challenge~\cite{agapie2016staying, eysenbach2005law}. Attrition, where people weaken or give up on their behavior change regimen, can be caused by a number of factors including low perceived intervention effectiveness~\cite{kovacs2018rotating}, high perceived intervention difficulty~\cite{given1985prediction}, lack of motivation~\cite{alfonsson2016motivation}, or a mismatch between the system's interventions and user preferences~\cite{kovacs2018rotating}.

In this paper, we seek to understand how people maintain or weaken their behavior change regimens over long periods of time. Are we able to maintain  the interventions that we set in place? If we lose the battle, does it happen slowly or suddenly? And once we lose, do we resume our attempt or give up permanently? These questions are critical to the design of behavior change systems, as users may succumb to present-biased choices that are not in line with their long-term goals~\cite{loewenstein1992choice, shu2010procrastination}. In addition to the opportunity to inductively build theory around these questions, there is also a set of practical questions that this research answers: behavior change systems must decide on an appropriate difficulty level~\cite{collins2004conceptual, almirall2018developing}: too light a touch, and users might not change their behavior~\cite{racey2016systematic}, while excessively aggressive interventions may backfire~\cite{given1985prediction,michael2009progressive, anderson2004structured}. Knowledge of how user preferences vary over time can help a system identify an appropriate difficulty level for the user in the present moment~\cite{baker2012optimal, adams2013adaptive}. 

In this paper, we study how productivity intervention difficulty preferences change over time, and explore the tradeoffs in terms of time, attrition, and accuracy of asking users about difficulty preferences at various frequencies. We do so by running three studies on the HabitLab platform, an in-the-wild behavior change platform for helping users reduce their time spent online. 



An important first question is how users' intervention difficulty preferences evolve over time: what happens to the difficulty levels that users choose over time? How effectively do they stick to their original intended regimen? So, our first study observationally tracks changes in users' choices of intervention difficulty over time. We observe users initially choosing more difficult interventions, and later choosing easier ones, with over half of users eventually keeping the system installed but choosing to have no interventions at all. This result makes clear that user preferences are not static, meaning that any system would need to track changes over time, for example through prompts asking users about their preferences. 

Of course, prompting users has attentional and time costs leading to attrition if done too frequently~\cite{scollon2009experience}. Thus, in our second study, we observe the costs of prompts in terms of time spent and attrition rates, by randomizing the frequency at which we ask users to choose their desired intervention difficulty levels. We find that excessive prompting significantly increases attrition rates, but that occasional prompting is actually beneficial for retention. 

In our third study, we  investigate users' future intentions. Specifically, we allow users to not only weaken their interventions, but to save that preference for an hour, a day, or a week. While the most popular intervention level continues to be ``No Intervention'', the most popular request is to ask again immediately on the next visit. This combination recurs repeatedly, with users continually disabling the system for the current visit but requesting that it try again next time rather than stop asking. This snooze button behavior suggests that users remain optimistic about their potential for future behavior change, even if in practice it never materializes.

This paper contributes an analysis of changes in user intervention difficulty preferences in the context of online productivity. We find that users' hope springs eternal: while users choose easier intervention difficulty levels over time and short-term choices can be detrimental to their ability to save time, most users choose to have choices, seemingly expecting to return to more difficult interventions in the near future. 

%% file: relatedwork.tex
\section{Related Work}






\subsection{Self-Interruptions and Productivity Interventions}

Self-interruptions~\cite{jin2009self} are a widespread occurrence in the workplace~\cite{lim2012cyberloafing} and among students~\cite{mark2014stress}, which are characterized by users interrupting their work with social media~\cite{meier2016facebocrastination}, email~\cite{mark2016email}, and recreational web browsing~\cite{vitak2011personal}. The relationship of self-interruptions and social media use with well-being is complicated -- there can be benefits~\cite{burke2016relationship, reinecke2014entertainment, toma2013self, quoquab2015does}, but excessive social media use can also lead to reduced well-being~\cite{cheng2019understanding, burke2020social, kross2013facebook, tromholt2016facebook}.


A number of sociotechnical approaches have emerged to reduce self-interruptions, including deactivating social network accounts~\cite{baumer2013limiting}, internet addiction bootcamps~\cite{kwee2010treatment}, workplace site filters~\cite{glassman2015monitor}, time trackers~\cite{kim2017omnitrack, kim2016timeaware}, as well as various productivity interventions delivered via browser extensions~\cite{lyngs2019self, lyngs2020just, agapie2016staying, kovacs2018rotating}, phone applications~\cite{kim2017technology, kovacs2019conservation}, and chatbots~\cite{tseng2019overcoming}.

A challenge in the design of these systems is how much control to give users. In the case of workplace site filters, overly restrictive policies can lower productivity and employee satisfaction~\cite{de2006current}. However, if productivity interventions are controlled by users, they can easily be uninstalled or bypassed, and rely on the user remaining committed to continue using them~\cite{kovacs2018rotating, agapie2016staying}. Thus, productivity tools need to adapt to users~\cite{lyngs2020just}, which they could do by asking users about their intervention preferences and adapting interventions accordingly.

\subsection{Why Lapses Occur: Initial Expectations, Present-Biased Choices, and Self-Control}

Sometimes, users choose an intervention -- such as deactivating their Facebook account and pledging to never use it again -- only to give up and reactivate weeks later~\cite{baumer2013limiting}. User behavior can lapse for numerous reasons, including declining motivation~\cite{marlatt199710, prochaska1997transtheoretical, bouton2000learning}. Relapse management techniques, which aim to combat such lapses~\cite{marlatt1996taxonomy, middleton2013long, bouton2014behavior, schwarzer2008modeling, marx1982relapse}, are implemented by some behavior change systems. Examples include ``cheat points'', which allow temporary deviations from goals ~\cite{agapie2016staying}, or ``streak freezes'', which allow users to maintain a streak without performing the target behavior~\cite{huynh2017analysis}. Some studies allow users to choose their own intervention~\cite{schueller2010preferences, rosenzweig2019choose, paredes2014poptherapy}, but this paper is the first system to study changes in user intervention preferences as reflected by repeated intervention choices over time.


Do users have difficulty sticking to their behavior change regimens because they have unrealistic initial expectations? In dieting contexts, users tend to overestimate their self-control abilities and have unrealistic expectations of their ability to lose weight~\cite{sparks1995perceived}, though this varies by individual~\cite{davidson1997optimism}. 
Users likewise underestimate the amount of time they spend on email and instant messaging when using laptops during lectures~\cite{kraushaar2019examining}. However, while users underestimate the number of times they visit Facebook, they overestimate the time they spend on Facebook~\cite{ernala2020well, junco2013comparing} and online~\cite{scharkow2016accuracy, araujo2017much}.

Another reason why users struggle to achieve their behavior change goals is that users make short-term choices that conflict with their long-term goals~\cite{ajzen1991theory}. These manifest themselves as inability to delay gratification, lack of self-control, procrastination, and addiction~\cite{loewenstein1992choice, shu2010procrastination}. Present-biased choices can be attributed to a number of factors -- firstly, short-term benefits are more immediate and salient than long-term losses, leading us to discount future outcomes~\cite{ainslie2001breakdown, loewenstein1992choice, soman2005psychology}. 
Additionally, we are often certain of short-term benefits, while long-term effects are less certain, so we discount the uncertain, long-term outcomes~\cite{payne2013life}, or end up considering only a desirable subset of possible outcomes~\cite{shu2008future, koehler1991explanation}. Optimism can also play a role in present-biased choices, as we are often overly optimistic that we will not suffer from possible negative long-term consequences~\cite{kahneman1993timid, zauberman2005resource}. Self-control -- the ability to resist desires when they conflict with goals -- is a key predictor of success~\cite{duckworth2016stitch, duckworth2016stitch}. While self-control abilities vary between individuals, situational factors can also influence self-control in the moment~\cite{hofmann2012everyday}. Self-control theories have been used for designing better systems to combat distractions~\cite{lyngs2019self}.

\subsection{Attrition in Behavior Change Systems}

Attrition is a major problem faced by behavior change systems~\cite{eysenbach2005law}. Within the HabitLab system, mismatches between users' intervention difficulty preferences and interventions shown by the system are commonly reported as a reason for uninstalling~\cite{kovacs2018rotating}, which motivates us to investigate adapting to user preferences as a means of potentially reducing attrition. That said, attrition in behavior change contexts is influenced by many factors, including lacking time~\cite{mcauley1990attrition}, motivation~\cite{burgess2017determinants}, enjoyment of interventions~\cite{van2019new}, the costs of interventions~\cite{iuga2014adherence}, lacking intention to change~\cite{belita2015attrition}, intervention novelty~\cite{kovacs2018rotating}, unintentionally forgetting about interventions~\cite{liu2013adherence}, or temporary lapses leading to abandonment~\cite{agapie2016staying}.

A number of technical approaches help address these issues -- for example, adaptive phone and email notifications can help remind users about interventions at the right time~\cite{kunzler2019context, linardon2020attrition}. Ambient interventions embedded into routinely used apps, smartwatches, homescreens, or lock screens can encourage engagement during downtime~\cite{kovacs2015feedlearn, zhao2018watch, consolvo2008activity, cai2015wait}. Gamification approaches such as streaks, points, and giving users cheat points can improve enjoyment and reduce abandonment after temporary lapses~\cite{alahaivala2016understanding, agapie2016staying, cugelman2013gamification, lyngs2019self, chou2019actionable}. Some systems ask users to make social and financial commitments to encourage them to stick to their goals~\cite{gine2010put, Panovich13habitbot}. Many systems for controlling time online or on phones show interventions automatically during usage, thus reducing attrition via defaults -- user inaction will not lead to attrition, as the tools need to be explicitly uninstalled~\cite{lyngs2019self, kovacs2018rotating, kovacs2019conservation, okeke2018good}. 

\subsection{Promoting Behavior Change}
There are several theoretical frameworks of behavior change~\cite{prochaska1997transtheoretical, bandura1977self, rosen2000sequencing, michie2005making, fogg2009behavior, ryan2008facilitating, teixeira2020classification, rhodes2004differentiating, ajzen1991theory}. Many of these theories put focal emphasis on the user's commitment to the behavior change regimen~\cite{fogg2009behavior, michie2005making, teixeira2020classification, rhodes2004differentiating}. Fogg's B=MAT model, for example, considers behavior change to occur in the presence of motivation, the ability to take action, and a trigger that prompts people to take action~\cite{fogg2009behavior}. However, these levels of commitment are challenging to measure, as they depend on the behavior change domain and numerous factors~\cite{martin2001student, artelt2005cross}, so most approaches rely on self-reporting, which may be unreliable~\cite{fulmer2009review}.
Some behavior change techniques can provide measurements on related proxies instead~\cite{oduor2017commitment, fulmer2009review, yeung2004octagon}. Commitment devices are arrangements where people commit to a plan for achieving a certain behavior goal in the future~\cite{bryan2010commitment}. They encourage people to stick to their goals by making commitments, such as financial~\cite{gine2010put, halpern2012commitment} or social~\cite{gugerty2007you} commitments. In our work, we draw on the theory of self-shaping: installing software to show interventions, and choosing intervention difficulty levels, can be thought of as a self-shaping commitment device~\cite{moraveji2011role}.



The field of behavioral economics has developed a number of theoretical frameworks for how to present choices to influence people's choices, known as choice architectures~\cite{thaler1980toward, johnson2012beyond}. Defaults are a well-known choice architecture which work by exploiting the status-quo bias~\cite{samuelson1988status}. Other widely used choice architectures include limiting the number of choices~\cite{cronqvist2004design}, sorting choices~\cite{lynch2000wine}, grouping choices~\cite{fox2005subjective}, and simplifying choice attributes to be more easily interpretable~\cite{peters2009bringing, soll2013consumer}. 
A number of choice architectures have been developed to combat our bias towards present-biased choices, aversion to uncertainty, and lead us to choices that have better long-term outcomes~\cite{johnson2012beyond, weber2007asymmetric, soman2005psychology}. 

Existing studies on self-control and choice architectures have studied contexts where choices only need to be made once or infrequently, and feedback and measurements are often delayed~\cite{johnson2012beyond}. The context of online productivity provides a superb domain for studying changing user preferences and choices, as we can prompt users multiple times per day, we can vary the frequency of prompting, and the system can provide immediate feedback in response to user choices~\cite{kovacs2018rotating}. This provides us with a more fine-grained lens on how users' preferences change over time.



%% file: researchquestions.tex

\subsection{Changing Preferences Over Time in Behavior Change Systems}
Prior work demonstrates that users will struggle to adhere to their behavior change goals, but the temporal dynamics of this process remain unknown.
In this paper, we explore, if users are allowed to choose the difficulty of their interventions, how they navigate the tradeoffs inherent to managing ideal difficulty, and the tradeoffs of different strategies that a behavior change system can pursue to adapt to these changing preferences. This leads us to the following research questions:

\textit{RQ1: How do users' intervention difficulty choices change over time?} If users' intervention difficulty preferences do not change over time, then behavior change systems can just ask users their preferences during onboarding. However, if they change over time, then behavior change systems may need to continually adapt to users' changing intervention difficulty preferences.  

\textit{RQ2: Should a behavior change system ask users about their difficulty preferences, and if yes, when and how often?} If preferences shift, but the system remains with the user's initial difficulty setting, it could lead to friction or discontinued use. Should the system prompt users about their preferences --- or will the act of asking itself cause attrition? How often should systems prompt users --- while more frequent prompting may allow us to more accurately model users' preferences, excessive prompting may have time costs and lead to attrition. We will explore the tradeoffs of prompting frequency with regards to time costs, attrition, and prediction accuracy. 

\textit{RQ3: Do users prefer to be asked about their intervention difficulty preferences, and if yes, how often do they prefer to be asked?} If users' difficulty preferences do not frequently change, we would expect that users would choose to be asked about their difficulty preferences infrequently. However, if users choose to be frequently prompted about their difficulty preferences, yet they keep choosing the same difficulty, this might suggest that users are expecting their future choices to differ from their current choice.  


%% file: system.tex
\section{Experimental Platform: The HabitLab Behavior Change System}

\begin{figure*}
	\includegraphics[width=1.0\textwidth]{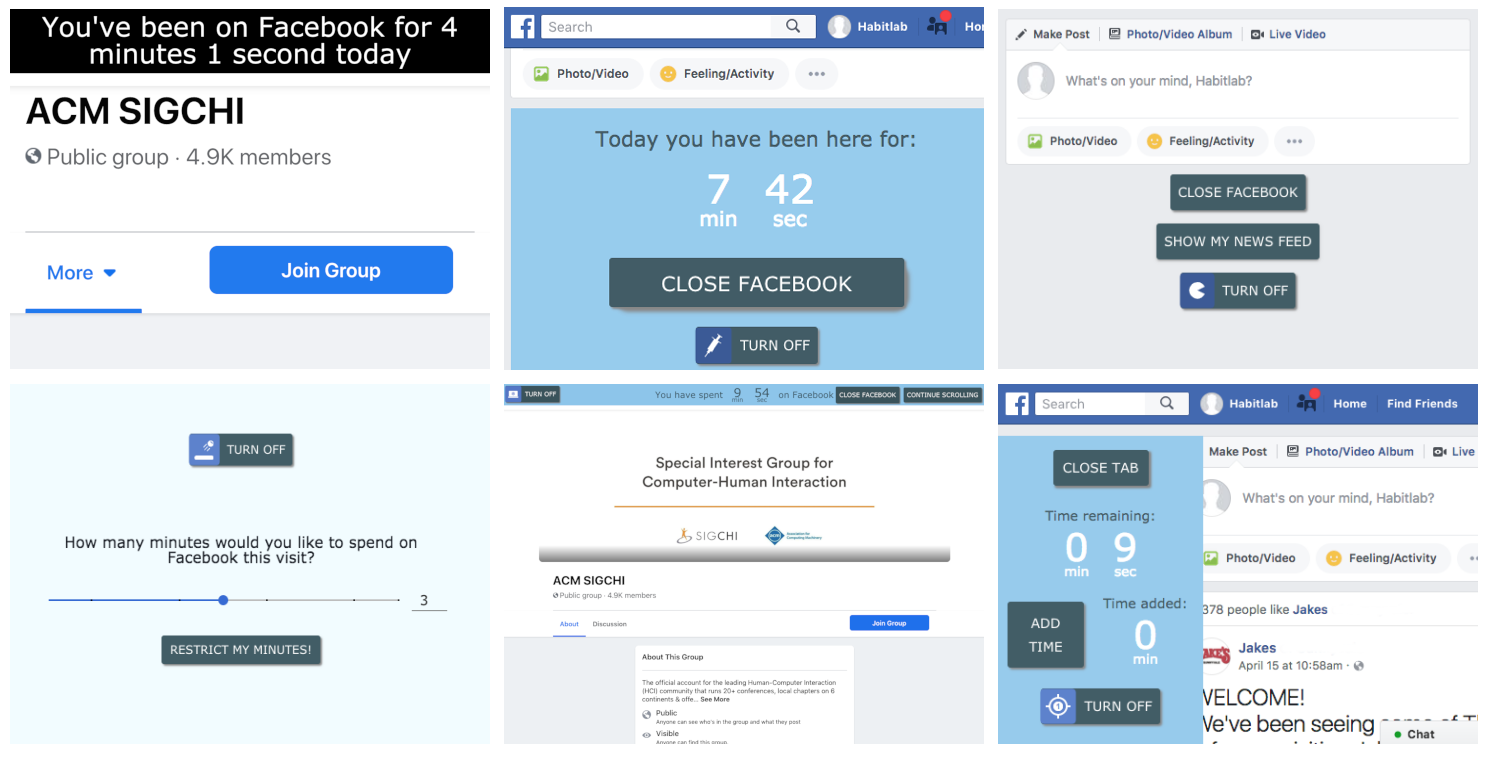}
	\caption{Examples of a few of the many HabitLab interventions available for reducing time on Facebook. From left to right, top to bottom: a timer showing time spent on site at the top of screen (difficulty rated as Easy); a timer injected into the news feed (Easy); requiring the user to opt-in to show the news feed (Medium); requiring the user to set a time limit for how long they will spend this session (Medium); preventing scrolling after a certain number of scrolls until the user clicks a button (Hard); a countdown timer that automatically closes the tab after time elapses (Hard)}
	\Description{Examples of a few of the many HabitLab interventions available for reducing time on Facebook. From left to right, top to bottom: a timer showing time spent on site at the top of screen (difficulty rated as Easy); a timer injected into the news feed (Easy); requiring the user to opt-in to show the news feed (Medium); requiring the user to set a time limit for how long they will spend this session (Medium); preventing scrolling after a certain number of scrolls until the user clicks a button (Hard); a countdown timer that automatically closes the tab after time elapses (Hard)}
	\label{fig:intervention_screenshots}
\end{figure*}


To answer these research questions, we conducted three studies on HabitLab~\cite{kovacs2018rotating, kovacs2019conservation}, an in-the-wild behavior change experimentation platform where users participate in behavior change experiments to help them reduce time online. Users install the browser extension, select sites they wish to reduce time on (\textit{goal sites}), and are shown various productivity interventions when they visit those sites, such as those shown in Figure~\ref{fig:intervention_screenshots}.\footnote{Descriptions of additional interventions can be found in Supplement A. More details about the HabitLab system can found in~\cite{kovacs2018rotating, kovacs2019conservation}.}

\subsection{Participant Demographics}

All participants of the studies in this paper were not recruited or compensated, but were rather all organic installs who discovered HabitLab though sources such as the website, the listing on the Chrome extension store, or press coverage in sources such as Wired or the New York Times. All users whose data we analyzed consented to participate in studies and share their data for research purposes upon installation.

As of this analysis, the HabitLab browser extension has over 12,000 daily active users. According to Google Analytics, 81\% of HabitLab users are male, and the most represented age group is 25 to 34. Users are from over 150 countries, and the most represented countries are the USA, Spain, Germany, and Russia. The goal sites they most commonly chose to reduce their time on were Facebook, Youtube, Twitter, Reddit, Gmail, Netflix, and VK. 

\subsection{Interventions and Difficulty Levels}

HabitLab includes interventions to help users reduce their time online, some of which are shown in Figure~\ref{fig:intervention_screenshots}. Some interventions are designed for specific websites such as Facebook, while others are generic and can be used on all sites.

We wished to categorize interventions into difficulty levels. We did so by asking three independent raters (HabitLab users who had been using the platform for over a month) to rate the difficulty level of each intervention as easy, medium, or hard. We opted for a 3-level difficulty categorization, as our studies ask users to choose difficulty levels and we did not want to overwhelm them with too many choices. We took the intervention's difficulty to be the median of its ratings. The Intraclass Correlation Coefficient~\cite{shrout1979intraclass}, a statistical measure of inter-rater agreement for ordinal data, was 0.53 -- indicating that intervention difficulty perceptions may vary between users. Intervention ratings, descriptions, and their effectiveness can be found in Supplement A. 




\begin{figure}
\centering
	\includegraphics[width=0.48\textwidth]{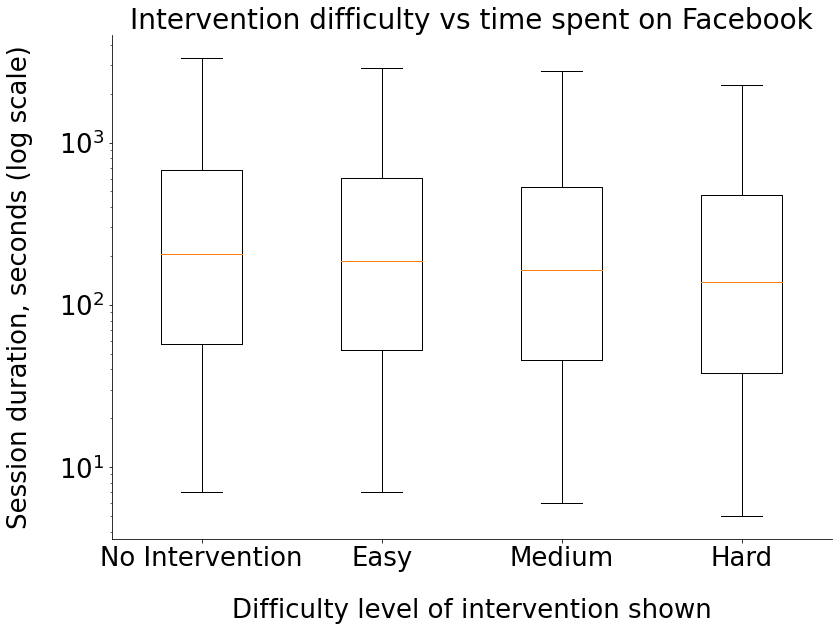}
	\caption{Box plot of Facebook session durations in the presence of interventions. Sessions are significantly shorter in the presence of more difficult interventions. Any intervention difficulty level is more effective than no intervention.} 
	\Description{Box plot of Facebook session durations in the presence of interventions. Sessions are significantly shorter in the presence of more difficult interventions. Any intervention difficulty level is more effective than no intervention.}
\label{fig:effectiveness_vs_difficulty}
\end{figure}

While our definition of intervention difficulty is based on difficulty ratings as opposed to observed effectiveness, interventions rated as more difficult are also more effective. We tested this in a study where on each visit to Facebook, a randomly chosen intervention (or no intervention) is shown. We then measure time spent on Facebook in the presence of that intervention. 

A total of 14,139,727 exposure samples were used in this study, from 14,834 users\footnote{A more detailed version of this study with Mann-Whitney U-statistic values and per-intervention analyses can be found in Supplement A.}. 
Our investigation revealed that the most time is spent when there was no intervention (median of 199 seconds per session), followed by easy (185 seconds), medium (161 seconds), and hard (135 seconds) interventions, as shown in Figure~\ref{fig:effectiveness_vs_difficulty}. There is a significant effect of difficulty on effectiveness according to a Kruskal-Wallis H test (H=37654, p < 0.001). Differences between pairs of groups are all statistically significant (p < 0.001) according to pairwise Mann-Whitney U tests. 
From this result, we conclude that the difficulty labels capture not only raters' opinions, but also are associated with monotonically increasing time savings when deployed, suggesting that they are in practice more effective. 




%% file: study1.tex
\begin{figure*}[hbt!]
\centering
	\includegraphics[width=0.73\textwidth]{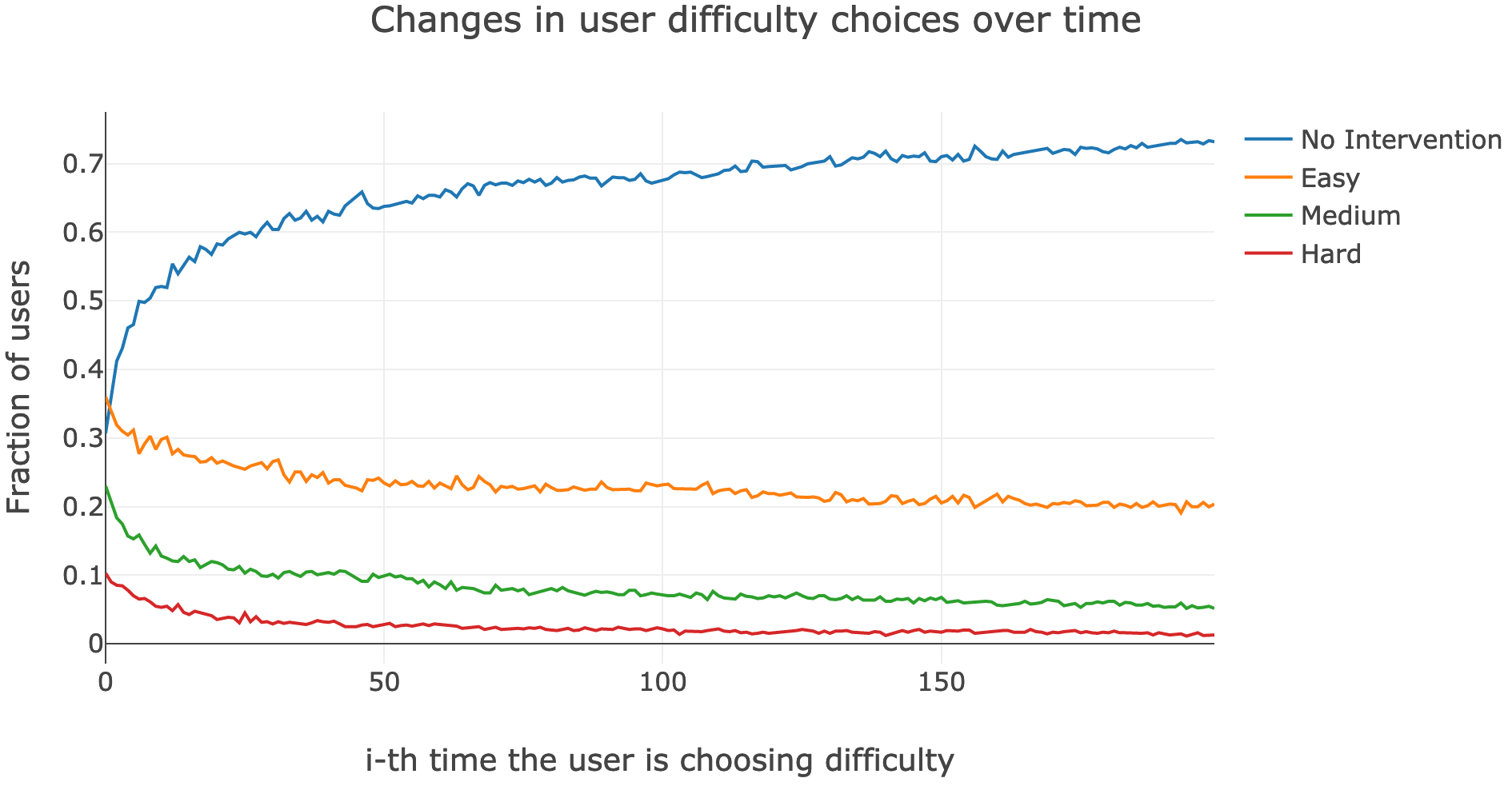}
	\caption{Changes in intervention difficulties chosen by users over time. Users gravitate towards easier interventions over time.} 
	\Description{Changes in intervention difficulties chosen by users over time. Users gravitate towards easier interventions over time.}
\label{fig:difficulty_over_time_lines}
\end{figure*}

\begin{figure}
    \centering
    \includegraphics[width=0.40\textwidth]{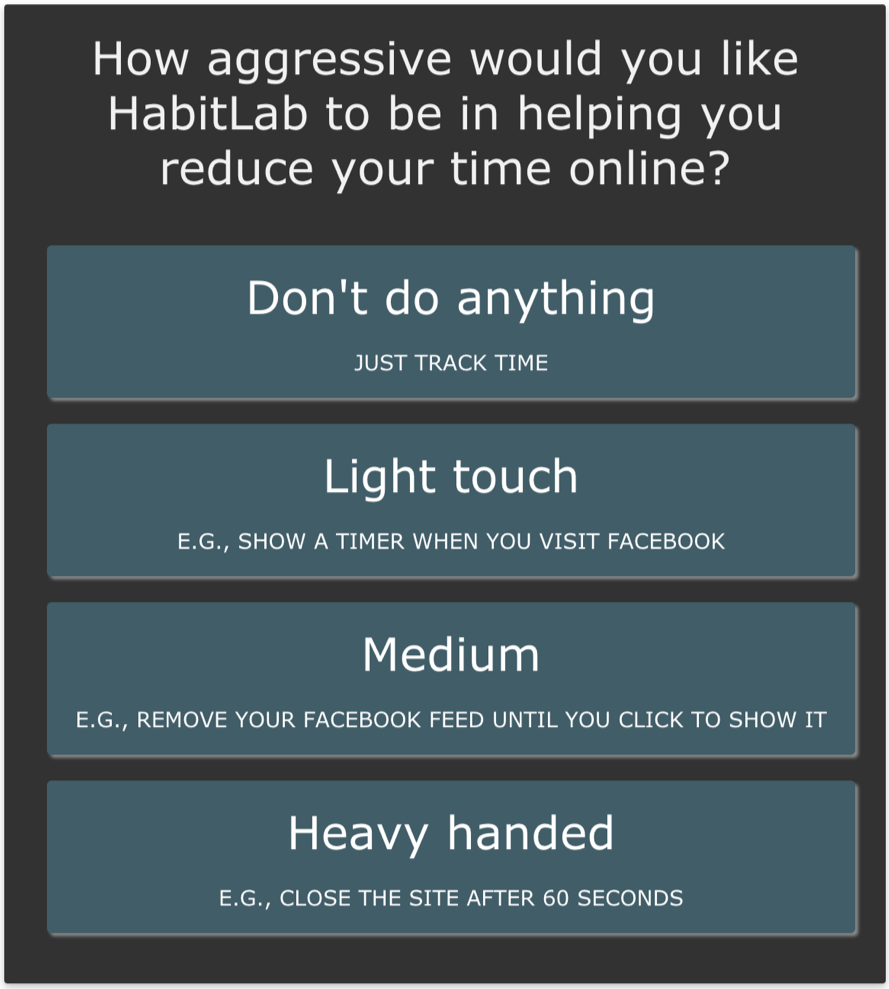}
    	\caption{The prompt through which we ask users their preferred intervention difficulty upon visiting a site. A similar prompt is also shown during onboarding.} 
    	\Description{The prompt through which we ask users their preferred intervention difficulty upon visiting a site. A similar prompt is also shown during onboarding.}
        \label{fig:difficulty_prompt}
\end{figure}

\section{Study 1: Changes in Intervention Difficulty Choices Over Time}


In our first study, we seek to understand temporal patterns in how users make choices that balance their commitment to their behavior change regimen against their interest in browsing a goal site. We do so by measuring how users' intervention difficulty preferences change over time, as observed through the intervention difficulty levels they choose on HabitLab. If these preferences are static, then we can just ask about preferences once, and keep them as-is. Many behavior change systems implicitly make this assumption, as they only ask the user to state their goals and configure the system during onboarding, and do not later revisit these goals to see whether the user's preferences have changed over time. If intervention difficulty preferences change over time, then understanding the trends will allow our systems to better tailor interventions to users. 

\subsection{Methodology}



When users install HabitLab, we prompt them during onboarding to choose how difficult they would like the interventions to be: No Intervention (``Don't do anything: just track time''), Easy (``Light touch''), Medium (``Medium''), or Hard (``Heavy handed''). Each option is annotated with an example intervention at that difficulty level. Later, as the user continues to use the system, we ask them via a periodic prompt on each visit to a goal site how difficult they would like to have their intervention for that visit, as shown in Figure~\ref{fig:difficulty_prompt}. By tracking changes in the chosen difficulty levels and how they differ from initial preferences indicated during onboarding, we can see how preferences change over time.

\begin{figure}
    \centering
    \includegraphics[width=0.48\textwidth]{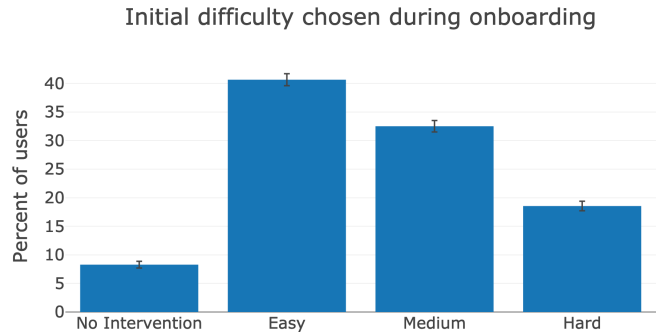}
    	\caption{Intervention difficulties chosen by users during onboarding. The most commonly chosen difficulty is easy interventions. Error bars indicate 95\% confidence intervals.} 
    	\Description{Intervention difficulties chosen by users during onboarding. The most commonly chosen difficulty is easy interventions. Error bars indicate 95\% confidence intervals.}
        \label{fig:difficulty_onboarding_percents}
\end{figure}

\begin{figure*}[hbt!]
\centering
	\includegraphics[width=1.0\textwidth]{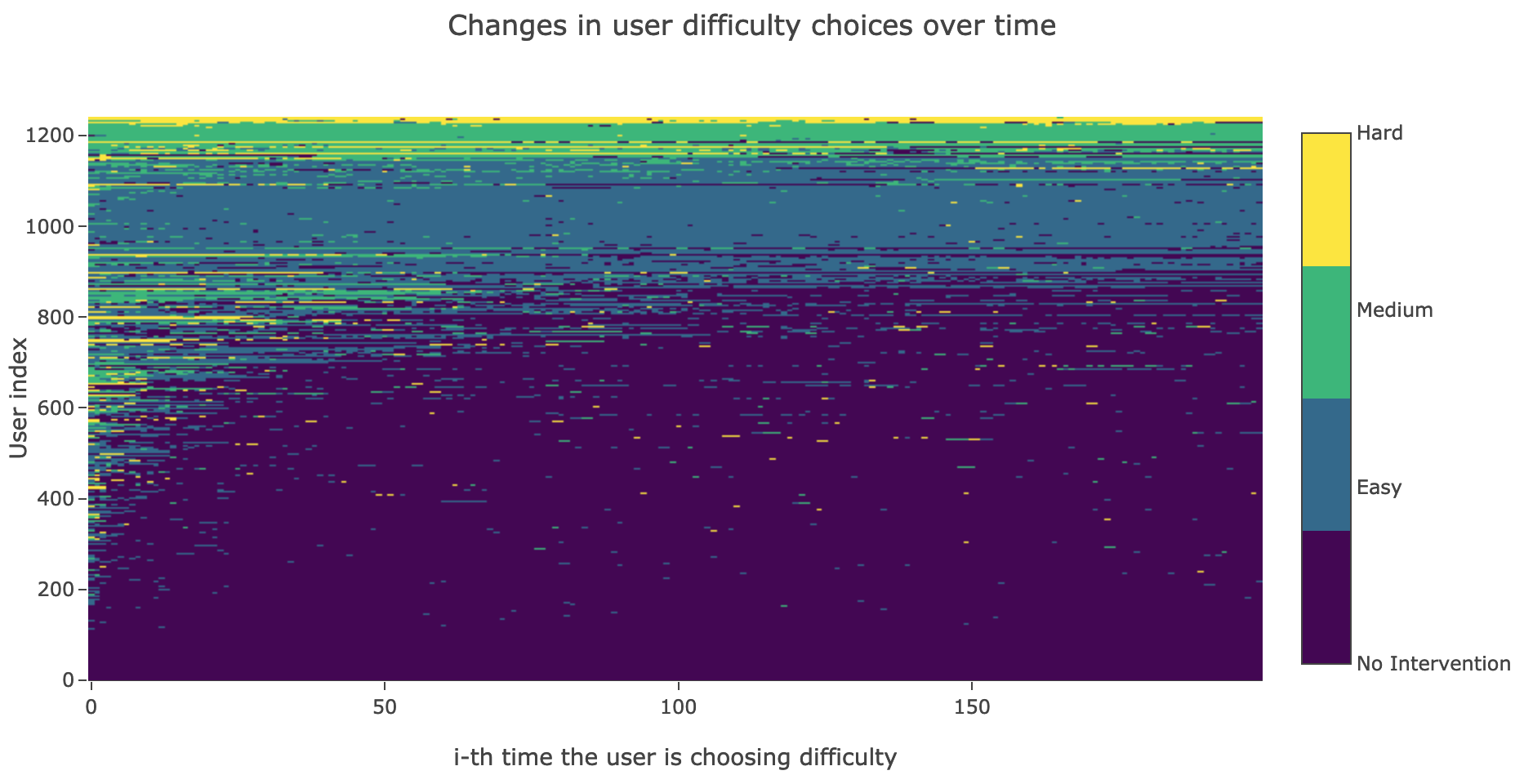}
	\caption{Users begin with more challenging interventions (left) but most slide to easy or no intervention over time (right). The first 200 intervention difficulty choices by each of the 1240 users who made at least 200 intervention difficulty choices. Each user is represented as a row. Time proceeds from left to right. The choice of difficulty is represented by color.} 
	\Description{Users begin with more challenging interventions (left) but most slide to easy or no intervention over time (right). The first 200 intervention difficulty choices by each of the 1240 users who made at least 200 intervention difficulty choices. Each user is represented as a row. Time proceeds from left to right. The choice of difficulty is represented by color.}
\label{fig:difficulty_over_time}
\end{figure*}

\subsection{Results}


Responses to the onboarding question of how difficult they would like to have their interventions are shown in Figure~\ref{fig:difficulty_onboarding_percents}. The question was answered by 8,372 users. The majority of users desire some form of intervention, with easy interventions being the most frequently chosen option, and no interventions being the least frequently chosen option. A chi-square test indicates there is a significant difference in proportions of responses ($\chi^2=2083.4, p<0.001$). Post-hoc tests on the resulting chi-square contingency table~\cite{fife2017package, sharpe2015chi} indicate that all pairs of differences are significant ($p<0.001$). 


As we are interested in changes in intervention difficulty preferences over time, we study successive responses to difficulty choice prompts over time. We consider users who have seen and selected an intervention difficulty at least 200 times: a total of 1,240 users during our study period. We visualize this in 2 figures: Figure~\ref{fig:difficulty_over_time_lines} shows the percent of users who choose each difficulty level at each of the 200 timesteps. Figure~\ref{fig:difficulty_over_time} visualizes each of the first 200 difficulty choices by each of the user 1,240 users. User preferences initially have a majority of users choosing to have interventions, and many initially go through an exploration phase where they try out different intervention difficulties, which can be seen in Figure~\ref{fig:difficulty_over_time} as changing colors on the left side. However, over time users choose progressively easier interventions, with 73\% of users choosing to be shown no intervention by their 200\textsuperscript{th} visit, as can be seen in Figure~\ref{fig:difficulty_over_time_lines}. 

%% file: study2.tex
\section{Study 2: Costs and Tradeoffs of Difficulty Choice Prompts}

In the first study, we showed that users' intervention difficulty preferences change over time, as indicated by their intervention difficulty choices. Thus, if a behavior change system aims to give users interventions of their desired difficulty, we cannot simply ask about preferences once during onboarding and assume they remains static -- the system must continually adapt. 

This situation creates challenges for system designers: continually asking users about their preferences may be burdensome and result in attrition. In this section, we measure the costs of asking users to choose a preferred intervention difficulty, and how frequently we need to  sample to be able to accurately predict the user's preferred difficulty choices. 

\begin{figure*}
\centering
	\includegraphics[width=1.0\textwidth]{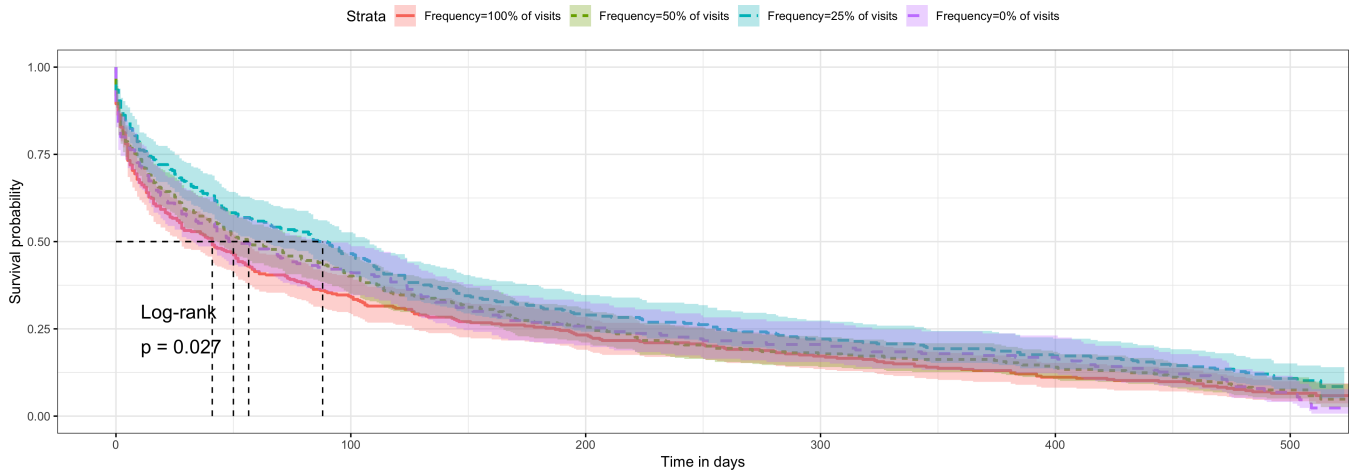}
	\caption{Effects of varying frequencies of difficulty choice prompts on retention rates. Retention is significantly higher when users are asked about their difficulty preferences at a low frequency (25\% of visits), compared to being asked on every visit.} 
	\Description{Effects of varying frequencies of difficulty choice prompts on retention rates. Retention is significantly higher when users are asked about their difficulty preferences at a low frequency (25\% of visits), compared to being asked on every visit.}
\label{fig:retention_costs}
\end{figure*}

\subsection{Time costs of difficulty choice prompts}

\label{section:time_costs}

\subsubsection{Methodology}

We can measure the time costs of difficulty choice prompts by observing the time it takes users to answer the difficulty prompt shown in Figure~\ref{fig:difficulty_prompt}. The time we measure is from when the prompt appears on screen, until the user selects a choice. We consider only sessions where the user actually answers the prompt, as opposed to simply ignoring it.

To determine whether showing the difficulty prompt results in a significant change in duration of visits to goal sites, each time a user visits a goal site the prompt is randomly shown with 50\% probability. We measure the overall session lengths -- that is, the total time spent from when the user visits a domain until they leave it -- when the difficulty prompt is shown, vs. when it is not shown. If the difficulty prompt is shown, an intervention of the chosen difficulty is shown; if the difficulty prompt is not shown, an intervention of the most recently chosen difficulty is shown. We then use a linear mixed model~\cite{nelder1972generalized} to compare log-normalized session lengths in sessions where the prompt was shown to sessions where it was not shown, controlling for the site and user as random effects. 

\subsubsection{Results}

\begin{figure}
    \centering
    \includegraphics[width=0.48\textwidth]{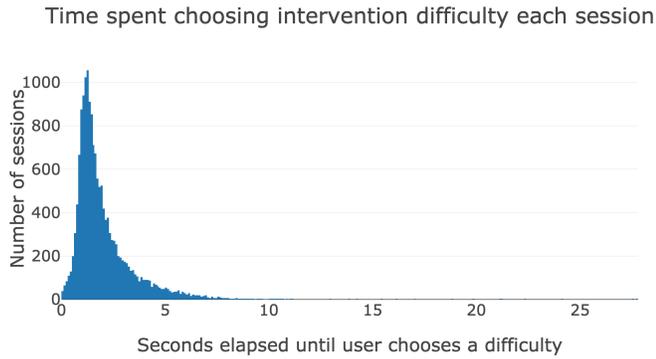}
	\caption{Histogram of time spent answering the difficulty choice prompt. Only sessions where the user made a choice are included. The median is 1.55 seconds.} 
	\Description{Histogram of time spent answering the difficulty choice prompt. Only sessions where the user made a choice are included. The median is 1.55 seconds.}
    \label{fig:time_costs}
\end{figure}

A histogram showing the time spent answering the difficulty prompt is shown in Figure~\ref{fig:time_costs}. This represents 16,183 responses from 1,831 users. The median of the distribution is 1.55 seconds. We find that there is no significant difference in the duration of sessions when an difficulty prompt is shown, vs not shown. Thus, from the perspective of time spent answering the prompt, our prompts for measuring user difficulty preferences appears to not have major costs. Additionally, the prompt itself does not appear to be influencing time spent on sites. 


\begin{table}
    \centering
\begin{tabular}{|l|l|l|l|}
\hline
Frequency & Beta (SE) & Hazard Ratio (95\% CI) & p\\
\hline
100\% of visits (ref) & - & - & -\\
\hline
50\% of visits & -0.09 (0.08) & 0.91 (0.78, 1.08) & 0.28\\
\hline
25\% of visits & -0.25 (0.09) & 0.78 (0.66, 0.92) & 0.003\\
\hline
0\% of visits & -0.07 (0.09) & 0.93 (0.78, 1.12) & 0.47\\
\hline
\end{tabular}

\vspace{0.5em}

\begin{tabular}{|l|l|}
\hline
Number of events & 1,029\\
\hline
Observations & 1,108\\
\hline
Concordance & 0.533 (SE = 0.01)\\
\hline
Likelihood ratio test & 9.43 (df=3, p=0.02)\\
\hline
Wald test & 9.24 (df=3, p=0.03)\\
\hline
Log rank test & 9.27 (df=3, p=0.03)\\
\hline
\end{tabular}
\vspace{4mm}
\caption{Effects of varying frequencies of difficulty choice prompts on retention rates. Retention is significantly higher when the prompts are shown 25\% of the time, compared to 100\% of the time (indicated by the hazard ratio).}
\Description{Effects of varying frequencies of difficulty choice prompts on retention rates. Retention is significantly higher when the prompts are shown 25\% of the time, compared to 100\% of the time (indicated by the hazard ratio).}
    \label{tab:retention_costs_table}
\end{table}

\subsection{Effects of difficulty choice prompt frequency on retention}

\label{section:retention_costs}

The costs of difficulty choice prompts are not restricted to time -- they may annoy and distract users, leading to attrition. We received feedback from many users that they had uninstalled HabitLab because they were annoyed by excessive difficulty choice prompts. That said, users often enjoy having their preferences taken into account, and difficulty choice prompts might help users gain a sense of control over the system. Hence, we hypothesized that there may be a tradeoff, with occasional prompting being beneficial, but excessive prompting increasing attrition.

\subsubsection{Methodology}

We performed an experiment in which we measure the causal effect of changing the interval of difficulty choice prompts on attrition. To do so, we randomly assign users into different conditions according to how frequently we show the difficulty choice prompt. There are 4 conditions: users can be asked 0\% of visits, 25\% of visits, 50\% of visits, or 100\% of visits. On visits where a prompt was shown, users are shown an intervention of the difficulty they choose in the prompt. If a prompt was not shown, users are shown an intervention of the difficulty they chose the last time they answered the prompt (falling back their difficulty choice during onboarding if they have not yet seen any prompts).\footnote{We ran a similar experiment where conditions were: each visit / daily / user-chosen intervals, and found similar results; see Supplement C.} We ran this experiment with 1108 users over 528 days, and analyzed retention using a Cox hazard regression model~\cite{cox1972regression}.

\subsubsection{Results}

The Cox hazard regression model showing user retention in the different experiment conditions is visualized in Figure~\ref{fig:retention_costs}, while the numeric results are shown in Table~\ref{tab:retention_costs_table}. We observe via the log-rank test in Table~\ref{tab:retention_costs_table} that there is a significant effect ($p<0.05$) of the prompting frequency on retention. The hazard ratio between the conditions where users are shown the difficulty prompt 25\% of the time, and 100\% of the time, is below 1 (0.78, see Table~\ref{tab:retention_costs_table}) and this difference is statistically significant ($p<0.005$). This means that user retention is significantly higher if users are asked the difficulty question at low frequency (25\% of visits), compared to on all visits. We believe it is because showing the prompt occasionally may be beneficial in terms of giving the user a reminder of the system's presence and granting the user a sense of control. This suggests that asking users about their difficulty preferences at low frequency can strike the right balance of giving users control, while not excessively annoying them.

\begin{figure}
    \centering
    \includegraphics[width=0.48\textwidth]{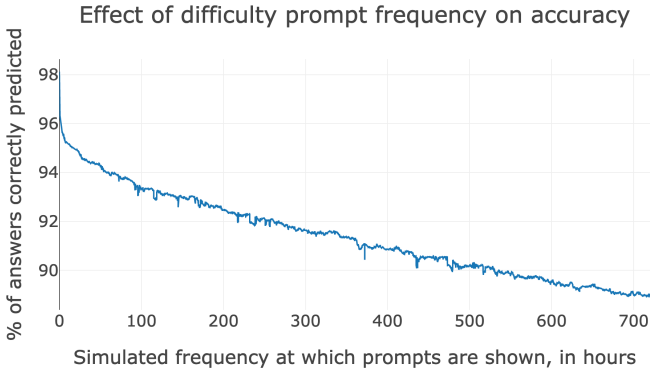}
	\caption{Accuracy of predicting users' intervention difficulty preferences, given different simulated frequencies of difficulty choice prompts.} 
	\Description{Accuracy of predicting user responses, given different simulated frequencies of difficulty choice prompts.}
	\label{fig:frequency_accuracy}
\end{figure}

\subsection{Effects of difficulty choice prompt frequency on accuracy}

\label{section:frequency_accuracy}

Given that we found that asking users about their difficulty preferences every visit is detrimental to retention, and that asking the same question with lower frequency has higher retention, we would ideally like to use low-frequency difficulty choice prompts to model user preferences. However, there is a tradeoff between frequency and accuracy -- asking with lower frequency may lead to lower accuracy. Hence, in this analysis we simulate different frequencies of asking users for their difficulty preference, and observe the accuracy of predicting the actual difficulty chosen by the user. We show results for a model that will predict that the user will choose the same difficulty that they chose the last time they saw the difficulty choice prompt. As shown in Figure~\ref{fig:frequency_accuracy}, we can correctly predict the user's difficulty choice with 96.1\% accuracy if we show the difficulty choice prompt at most once per hour, 94.7\% accuracy if we ask at most once per day, or 92.8\% accuracy if we ask at most once per week. Hence, low frequency prompts are sufficient to accurately predict user intervention difficulty preferences. 

\section{Study 3: User Preferences for Difficulty Choice Prompts}

\label{section:frequency_preferences}

Our previous study measured behavioral outcomes such as attrition with various frequencies of difficulty choice prompts, but did not take into account user preference at all. We found that retention was improved by asking users about their preferred difficulty at a low frequency, yet there were some users who were sufficiently annoyed by difficulty choice prompts that it led them to uninstall. If given control over prompting frequency, would users want to slide to longer and longer windows of low difficulty or no interventions -- as would be consistent with the tendency to regress in difficulty we observed in Study 1? Or would they still choose low-frequency difficulty choice prompts, to maintain their sense of control over intervention difficulty? 


\begin{figure}
    \centering
    	\includegraphics[width=0.30\textwidth]{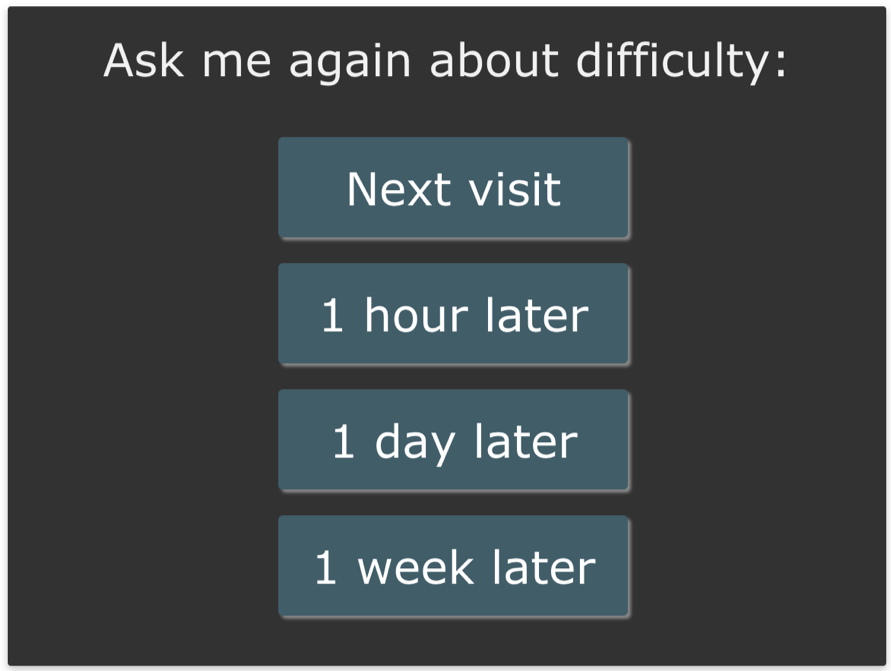}
    	\caption{Prompt asking users to choose when to be asked again about difficulty, shown after they choose a difficulty.}
    	\Description{Prompt asking users to choose when to be asked again about difficulty, shown after they choose a difficulty.}
    	\label{fig:frequency_prompt}
\end{figure}

\begin{figure*}[hbt!]
    \centering
	\includegraphics[width=1.0\textwidth]{figures/frequency_preferences_with_difficulty.png}
	\caption{Choices for intervention difficulty intersected with when to ask again about difficulty. The most commonly chosen option is to have no intervention this visit, but be asked again on the next visit. Error bars indicate 95\% confidence intervals.} 
	\Description{Choices for intervention difficulty intersected with when to ask again about difficulty. The most commonly chosen option is to have no intervention this visit, but be asked again on the next visit. Error bars indicate 95\% confidence intervals.}
    \label{fig:frequency_preferences_with_difficulty}
\end{figure*}

\subsection{Methodology}





In order to gather this data, we introduced an additional prompt shown immediately after the user selects the intervention difficulty which asks the user when they wish to be prompted again (Figure~\ref{fig:frequency_prompt}). The interventions that users are shown remain at the difficulty they chose until the next time the prompt is shown. We gathered 31,979 exposure samples from 644 users over the course of 385 days.

\begin{figure}
    \centering
    	\includegraphics[width=0.48\textwidth]{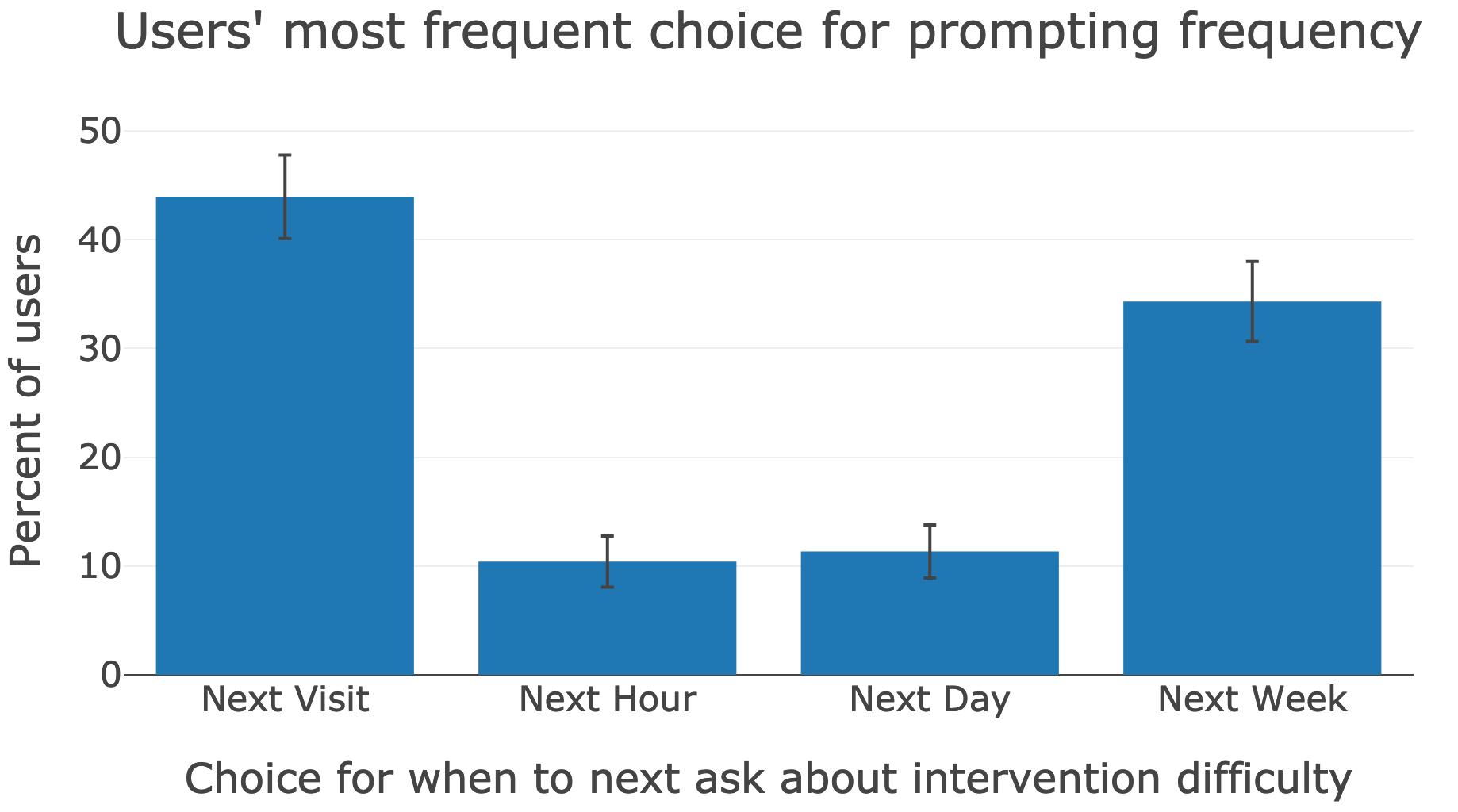}
	\caption{User preferences for frequency of difficulty choice prompts. A plurality (44\%) of users most commonly choose to be asked about intervention difficulty on the next visit. Error bars indicate 95\% confidence intervals.}
	\Description{User preferences for frequency of difficulty choice prompts. A plurality (44\%) of users most commonly choose to be asked about intervention difficulty on the next visit. Error bars indicate 95\% confidence intervals.}
	\label{fig:frequency_preferences}
\end{figure}

\begin{figure}
    \centering
	\includegraphics[width=0.48\textwidth]{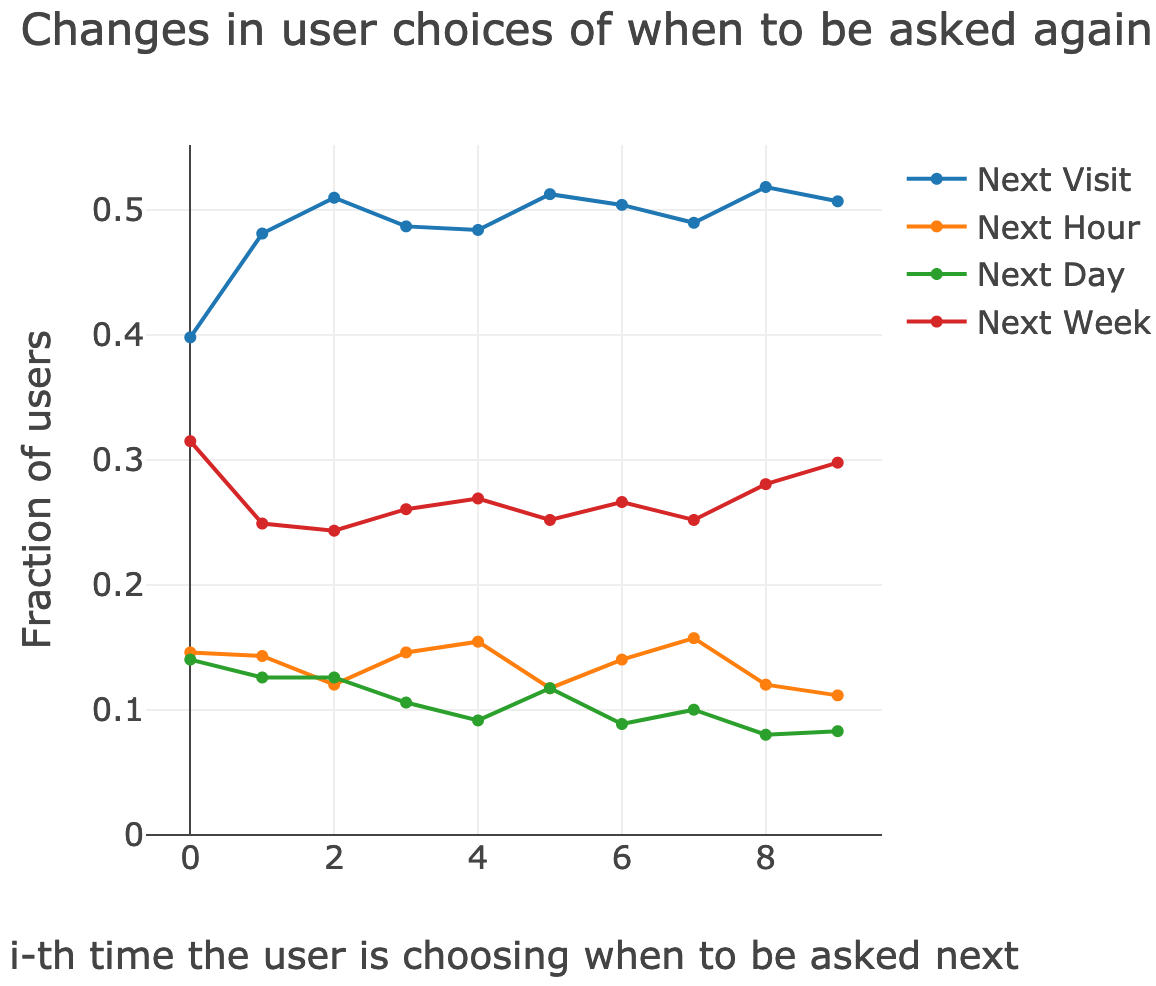}
	\caption{The first 10 choices of when to be prompted again, among the 349 users who made at least 10 choices.} 
	\Description{The first 10 choices of when to be prompted again, among the 349 users who made at least 10 choices.}
    \label{fig:frequency_preferences_over_time_lines_10}
\end{figure}

\subsection{Results}


User preferences for when they wish to be asked about intervention difficulty are shown in Figure~\ref{fig:frequency_preferences}. We observe that users most commonly choose the option to be asked again the next visit. A chi-square test indicates there is a significant difference in proportions of responses ($\chi^2=217.79, p<0.001$). Post-hoc tests on the resulting chi-square contingency table~\cite{fife2017package, sharpe2015chi} indicate that the difference between Next Hour vs Next Day is not significant ($p>0.5$), while all other pairs are significantly different ($p<0.01$ for Next Visit vs Next Week, and $p<0.001$ for all others).

To confirm that users' preferences to be asked again about difficulty the next visit is not just a transient phenomenon that goes away over time, we show the change in user choices over time, across the 349 users who made at least 10 choices, in Figure~\ref{fig:frequency_preferences_over_time_lines_10}. Here, we observe that users' choice of when to be asked again is mostly stable over time, and there is in fact a slight increase in the fraction of users choosing to be asked again next visit over the first 3 visits. This is the opposite trend of we would expect would result from fatigue due to excessive prompting -- which would be that users would choose to be shown prompts less frequently over time.\footnote{See Supplement D for additional visualizations and analyses over longer periods.}

We also visualize the intersection of the intervention difficulty chosen and when the user wishes to be asked again in Figure~\ref{fig:frequency_preferences_with_difficulty}. Here, we find that the most commonly chosen combination is to have no intervention this visit, but to be asked again in the following visit. This result is peculiar, as it seems irrational on the part of users. If users did not want to be bothered by interventions, the logical choice would be to show the intervention difficulty prompt as infrequently as possible -- that is, next week. If users wanted to change the intervention difficulty every visit, we would expect to see frequent changes from choosing no intervention to more difficult interventions. However, as we can see in Figure~\ref{fig:difficulty_over_time} this rarely happens -- once a user falls into a pattern of repeatedly choosing to have no intervention, they only occasionally deviate from it. 




%% file: discussion.tex
\section{Discussion}

This paper is motivated by our observation that users who uninstall HabitLab often cite a mismatch between the difficulty of interventions shown by HabitLab and the difficulty users would prefer as a reason for uninstalling~\cite{kovacs2018rotating}. As a result, we wish to understand changes in users' intervention difficulty preferences over time, and whether it is helpful for behavior change systems to adapt to users' difficulty preferences by prompting them to select intervention difficulty levels. We ran studies on the HabitLab platform to investigate the following research questions:

\textit{RQ1: How do users' intervention difficulty choices change over time?} We find that user choices of intervention difficulty decline over time. Thus, our behavior change system cannot simply ask users their preferences during onboarding and assume they will remain constant -- it needs to continually adapt to changing user preferences.

\textit{RQ2: Should a behavior change system ask users about their difficulty preferences, and if yes when and how often?} We find that prompting users for their intervention difficulty preference with low frequency has low time costs, and that low-frequency prompting reduces attrition compared to high-frequency prompting. Low-frequency prompting is sufficient to accurately predict user difficulty choices. Thus, a strategy of prompting at low frequency works well for both reducing attrition and adapting to user preferences.

\textit{RQ3: Do users prefer to be asked about their intervention difficulty preferences, and if yes, how often do they prefer to be asked?} If given a choice of when to be asked again, users will most commonly choose to have no intervention this visit, but to be asked again the next visit. Thus, given users continually ask to be asked again, they appear to expect their future intervention difficulty preferences to change, and do not mind being prompted.



Users are initially optimistic when choosing behavior change interventions -- perhaps unrealistically so.
We have found that users choose higher difficulty interventions during onboarding than they choose long-term. If we ask users about their desired intervention difficulty later on, it will progressively decline over time.

What are the factors underlying changes in users' intervention difficulty preferences over time? A decline in motivation might occur among HabitLab users over time -- in contexts such as volunteering, declines in motivation have been cited as possible reasons for people volunteering less over time~\cite{yanay2008decline}. However there are alternative explanations for changing intervention difficulty preferences -- it could indicate a shift in priorities, or a decline in commitment to the goal of reducing time online -- perhaps the user installed HabitLab to help them focus during a deadline, and once the deadline has passed they care less about reducing time online. Another alternative is that after repeated exposure, the effectiveness of the intervention declines~\cite{kovacs2018rotating}, and users may end up opting for no intervention because they find interventions more distracting than helpful. 

Despite users' tendency to choose easier interventions over time, they cling on to hope that they will get back on track. If we ask them when they wish to be asked again about intervention difficulty, by far the most common choice is to have no intervention this visit, but to ask again the next visit.

One explanation for these phenomenon is a combination of users focusing on immediate outcomes when making choices, but attempting to preserve their self-image to be in accordance with their long-term goals when planning about the future. While this behavior seems contradictory, it can be explained from the perspective of reducing cognitive dissonance~\cite{festinger1962theory}. Users wish to enjoy the short-term benefits of violating their long-term goals, yet still convince themselves that they will later return to pursuing their long-term goal so they avoid the feeling of having given up. Thus, just like a dieter confronted with temptation may promise themselves that they will only lapse this one time and will stick to their diet in the future --- only to repeatedly lapse in the future --- HabitLab users may convince themselves that they are only taking a break this one time, and will resume interventions in the future. By retaining the option of resuming in the future, they can continue to reassure themselves that they have not given up on their goal. 

An alternative explanation for our result that users consistently choose to have no intervention, but want to be asked again, is that the difficulty choice prompts themselves serve as an intervention that some users feel they need. Perhaps users enjoy the reminder, or they enjoy the sense of choosing, or they think the prompt itself is effective at reducing their time online. As seen by the increase in attrition when users are shown difficulty choice prompts at higher frequencies (Section~\ref{section:retention_costs}), and complaints we received from users about excessive prompting while running this study, this enjoyment of prompting likely does not apply to all users. Additionally, while the presence of prompting can change behavior~\cite{barrett2001introduction}, we did not observe a difference in time spent when the difficulty choice prompt was shown, vs when it was not (Section~\ref{section:time_costs}), so the prompts themselves do not seem effective as an intervention in this context. However, it is possible that despite annoying some users to the point that they uninstall, other users believe the difficulty choice prompt provides enough value that they constantly ask for it to be shown again.

If we decide we should ask users about their intervention preferences, how often should we do so? In our experiments studying the effects of varying prompting frequency on retention, we found that low prompting frequency results in the highest retention (Section~\ref{section:retention_costs} and Supplement C). Yet, when choosing when to be asked next about intervention difficulty -- which effectively allows users to choose their prompting frequency -- they most often choose to be asked again next visit (Section~\ref{section:frequency_preferences}), effectively choosing the highest prompting frequency. These results are not necessarily contradictory -- people often do not correctly predict their own long-term preferences~\cite{loewenstein1999wouldn, simonson1990effect}, often prefer retaining the ability to choose~\cite{leotti2010born}, the act of choosing itself can influence their future perceptions~\cite{luettgau2020decisions}, and there can be individual differences in tolerance for prompting~\cite{csikszentmihalyi2014validity}. As for frequency of prompting, if allowing users to choose themselves they split between the two extremes of low and high frequency (Section ~\ref{section:frequency_preferences}), and preferences do not change over time (Supplement D), but giving users additional control over prompting frequency does not appear to be beneficial for retention or effectiveness (Supplement B and C). This suggests that continually asking users about prompting frequency is not necessary, and keeping prompting at low frequency should be sufficient to avoid attrition.


How might we design behavior change systems for users who wish to retain their self-image as sticking to their long-term goals, but are continually tempted to break them via present-biased choices? It may be the case that the act of choosing itself impairs self-control~\cite{vohs2014making}, in which case one option is to remove choices entirely, and not tempt users. For example, just like we could prevent the dieter from encountering desserts and default them to salads, we can simply default users to harder interventions\footnote{We ran a study testing effects of removing choices and assigning users to default difficulties; see Supplement B.}. So long as the interventions are not so far above the user's preferred difficulty level that they drop out, they may continue to stick to it and enjoy the benefits.

Our observation that users weaken their intervention difficulty over time, but appear to hope to re-strengthen it soon, suggests that commitment devices~\cite{bryan2010commitment, michie2013behavior, barata2017promote, martin2012commitments} might be worth exploring as potential strategies for keeping users engaged with interventions. Platforms could capitalize on opportunities to try and get users to commit to a slightly more difficult set of interventions, knowing that it might weaken again later. For example, rather than asking the user what difficulty of intervention they would like this visit, we could ask them upfront what difficulty level they will want in the future. By removing the choice, the user's decision is less susceptible to influence from the present. Thus, user choice is not necessarily detrimental --- rather, choices should be designed in a way that steers users towards achieving their goals. 


\subsection{Limitations}



Given that perceptions of intervention difficulty may vary depending on the user, one may ask why we chose to have a single categorization of intervention difficulty levels, as opposed to developing a personalized categorization for each user. The latter approach would require users to try out and rate difficulties of all interventions during onboarding before they can start using the system -- a lengthy task that would result in our uncompensated, voluntary users not completing onboarding and instead uninstalling. Furthermore, a user's perception of intervention difficulty might change over time, so for truly accurate per-user intervention difficulty ratings we would need additional prompts, which would increase attrition. Per-user intervention difficulties would also complicate statistical analyses, as it would introduce per-user variation in the distributions of randomly chosen easy/medium/hard interventions.

One might ask how the difficulty levels chosen by users may relate to other measures, such as how motivated users are to save time online. Although it is possible that motivation may influence users' choices, it is not necessarily directly observable through users' intervention difficulty choices. We chose to measure users' difficulty choices rather than asking them about ``motivation'', as ``motivation'' is challenging to define and measure directly -- users may not accurately self-report motivation when asked~\cite{fulmer2009review}. 


Our methodology of asking users for their preferred intervention difficulty this visit, and when they would like to be asked again, is somewhat similar to commitment devices as well as experience sampling, but does not match the traditional definition of either. Commitment devices generally ask users to make future commitments~\cite{bryan2010commitment}. For example, ``Would you like to commit to hard interventions for next week?'' would be an example of a commitment device in this context. Experience sampling, in turn, differs from our methodology as it generally does not react to the user's choice by immediately presenting an intervention~\cite{hektner2007experience}. For example, asking ``How difficult did you find that intervention?'', and not acting on the response, would be an example of experience sampling in this context.

%% file: conclusion.tex
\section{Conclusion}

Attrition is a major problem faced by behavior change systems~\cite{eysenbach2005law, kovacs2018rotating}, and users commonly report mismatches between the difficulty of interventions shown by the system and users' preferred difficulty as a reason for attrition~\cite{kovacs2018rotating}. In this paper we have explored how users' intervention difficulty preferences change over time, and how behavior change systems can adapt to them. 
Using prompts on the HabitLab platform, we find that users choose higher intervention difficulty during onboarding, but that their choice of intervention difficulty declines over time. We find that asking users their preferred intervention difficulty at low frequency can both accurately predict the user's preference for intervention difficulty, and can be done at low cost in terms of time and attrition rates. 

If we allow users to choose both their desired intervention difficulty as well as when they will be prompted next, they overwhelmingly choose to have no intervention this visit, but to be prompted again next visit. However, users continue to request the system to do nothing, but ask again the next visit, and rarely end up later choosing harder interventions. We believe their choice to have no intervention this visit is driven by present-biased decisions that discount future outcomes, while their choice to be prompted again is driven by a wish to avoid cognitive dissonance and a belief that they will soon get back on track towards achieving their behavior change goals.


Many HCI systems aim to empower users by predicting and understanding users' intentions and preferences, and following them. In the case of behavior change systems, empowering users to achieve their goals requires us to understand users' preferences, while taking into consideration that users may be overly optimistic when initially choosing their behavior change regimen, and may succumb to present-biased choices over time. 

%% file: main.bbl

\begin{thebibliography}{137}


\ifx \showCODEN    \undefined \def \showCODEN     #1{\unskip}     \fi
\ifx \showDOI      \undefined \def \showDOI       #1{#1}\fi
\ifx \showISBNx    \undefined \def \showISBNx     #1{\unskip}     \fi
\ifx \showISBNxiii \undefined \def \showISBNxiii  #1{\unskip}     \fi
\ifx \showISSN     \undefined \def \showISSN      #1{\unskip}     \fi
\ifx \showLCCN     \undefined \def \showLCCN      #1{\unskip}     \fi
\ifx \shownote     \undefined \def \shownote      #1{#1}          \fi
\ifx \showarticletitle \undefined \def \showarticletitle #1{#1}   \fi
\ifx \showURL      \undefined \def \showURL       {\relax}        \fi
\providecommand\bibfield[2]{#2}
\providecommand\bibinfo[2]{#2}
\providecommand\natexlab[1]{#1}
\providecommand\showeprint[2][]{arXiv:#2}

\bibitem[\protect\citeauthoryear{Adams, Sallis, Norman, Hovell, Hekler, and
  Perata}{Adams et~al\mbox{.}}{2013}]%
        {adams2013adaptive}
\bibfield{author}{\bibinfo{person}{Marc~A Adams}, \bibinfo{person}{James~F
  Sallis}, \bibinfo{person}{Gregory~J Norman}, \bibinfo{person}{Melbourne~F
  Hovell}, \bibinfo{person}{Eric~B Hekler}, {and} \bibinfo{person}{Elyse
  Perata}.} \bibinfo{year}{2013}\natexlab{}.
\newblock \showarticletitle{An adaptive physical activity intervention for
  overweight adults: a randomized controlled trial}.
\newblock \bibinfo{journal}{\emph{PloS one}} \bibinfo{volume}{8},
  \bibinfo{number}{12} (\bibinfo{year}{2013}), \bibinfo{pages}{e82901}.
\newblock


\bibitem[\protect\citeauthoryear{Agapie, Avrahami, and Marlow}{Agapie
  et~al\mbox{.}}{2016}]%
        {agapie2016staying}
\bibfield{author}{\bibinfo{person}{Elena Agapie}, \bibinfo{person}{Daniel
  Avrahami}, {and} \bibinfo{person}{Jennifer Marlow}.}
  \bibinfo{year}{2016}\natexlab{}.
\newblock \showarticletitle{Staying the course: System-driven lapse management
  for supporting behavior change}. In \bibinfo{booktitle}{\emph{Proceedings of
  the 2016 CHI Conference on Human Factors in Computing Systems}}.
  \bibinfo{pages}{1072--1083}.
\newblock


\bibitem[\protect\citeauthoryear{Ainslie and George}{Ainslie and
  George}{2001}]%
        {ainslie2001breakdown}
\bibfield{author}{\bibinfo{person}{George Ainslie} {and}
  \bibinfo{person}{Ainslie George}.} \bibinfo{year}{2001}\natexlab{}.
\newblock \bibinfo{booktitle}{\emph{Breakdown of will}}.
\newblock \bibinfo{publisher}{Cambridge University Press}.
\newblock


\bibitem[\protect\citeauthoryear{Ajzen et~al\mbox{.}}{Ajzen
  et~al\mbox{.}}{1991}]%
        {ajzen1991theory}
\bibfield{author}{\bibinfo{person}{Icek Ajzen} {et~al\mbox{.}}}
  \bibinfo{year}{1991}\natexlab{}.
\newblock \showarticletitle{The theory of planned behavior}.
\newblock \bibinfo{journal}{\emph{Organizational behavior and human decision
  processes}} \bibinfo{volume}{50}, \bibinfo{number}{2} (\bibinfo{year}{1991}),
  \bibinfo{pages}{179--211}.
\newblock


\bibitem[\protect\citeauthoryear{Alah{\"a}iv{\"a}l{\"a} and
  Oinas-Kukkonen}{Alah{\"a}iv{\"a}l{\"a} and Oinas-Kukkonen}{2016}]%
        {alahaivala2016understanding}
\bibfield{author}{\bibinfo{person}{Tuomas Alah{\"a}iv{\"a}l{\"a}} {and}
  \bibinfo{person}{Harri Oinas-Kukkonen}.} \bibinfo{year}{2016}\natexlab{}.
\newblock \showarticletitle{Understanding persuasion contexts in health
  gamification: A systematic analysis of gamified health behavior change
  support systems literature}.
\newblock \bibinfo{journal}{\emph{International journal of medical
  informatics}}  \bibinfo{volume}{96} (\bibinfo{year}{2016}),
  \bibinfo{pages}{62--70}.
\newblock


\bibitem[\protect\citeauthoryear{Alfonsson, Olsson, and Hursti}{Alfonsson
  et~al\mbox{.}}{2016}]%
        {alfonsson2016motivation}
\bibfield{author}{\bibinfo{person}{Sven Alfonsson}, \bibinfo{person}{Erik
  Olsson}, {and} \bibinfo{person}{Timo Hursti}.}
  \bibinfo{year}{2016}\natexlab{}.
\newblock \showarticletitle{Motivation and treatment credibility predicts
  dropout, treatment adherence, and clinical outcomes in an internet-based
  cognitive behavioral relaxation program: a randomized controlled trial}.
\newblock \bibinfo{journal}{\emph{Journal of medical Internet research}}
  \bibinfo{volume}{18}, \bibinfo{number}{3} (\bibinfo{year}{2016}),
  \bibinfo{pages}{e52}.
\newblock


\bibitem[\protect\citeauthoryear{Almirall, Kasari, McCaffrey, and
  Nahum-Shani}{Almirall et~al\mbox{.}}{2018}]%
        {almirall2018developing}
\bibfield{author}{\bibinfo{person}{Daniel Almirall}, \bibinfo{person}{Connie
  Kasari}, \bibinfo{person}{Daniel~F McCaffrey}, {and} \bibinfo{person}{Inbal
  Nahum-Shani}.} \bibinfo{year}{2018}\natexlab{}.
\newblock \showarticletitle{Developing optimized adaptive interventions in
  education}.
\newblock \bibinfo{journal}{\emph{Journal of research on educational
  effectiveness}} \bibinfo{volume}{11}, \bibinfo{number}{1}
  (\bibinfo{year}{2018}), \bibinfo{pages}{27--34}.
\newblock


\bibitem[\protect\citeauthoryear{Anderson, Luan, and H{\o}ie}{Anderson
  et~al\mbox{.}}{2004}]%
        {anderson2004structured}
\bibfield{author}{\bibinfo{person}{James~W Anderson}, \bibinfo{person}{Jingyu
  Luan}, {and} \bibinfo{person}{Lars~H H{\o}ie}.}
  \bibinfo{year}{2004}\natexlab{}.
\newblock \showarticletitle{Structured weight-loss programs: meta-analysis of
  weight loss at 24 weeks and assessment of effects of intervention intensity}.
\newblock \bibinfo{journal}{\emph{Advances in Therapy}} \bibinfo{volume}{21},
  \bibinfo{number}{2} (\bibinfo{year}{2004}), \bibinfo{pages}{61--75}.
\newblock


\bibitem[\protect\citeauthoryear{Araujo, Wonneberger, Neijens, and
  de~Vreese}{Araujo et~al\mbox{.}}{2017}]%
        {araujo2017much}
\bibfield{author}{\bibinfo{person}{Theo Araujo}, \bibinfo{person}{Anke
  Wonneberger}, \bibinfo{person}{Peter Neijens}, {and} \bibinfo{person}{Claes
  de Vreese}.} \bibinfo{year}{2017}\natexlab{}.
\newblock \showarticletitle{How much time do you spend online? Understanding
  and improving the accuracy of self-reported measures of Internet use}.
\newblock \bibinfo{journal}{\emph{Communication Methods and Measures}}
  \bibinfo{volume}{11}, \bibinfo{number}{3} (\bibinfo{year}{2017}),
  \bibinfo{pages}{173--190}.
\newblock


\bibitem[\protect\citeauthoryear{Artelt}{Artelt}{2005}]%
        {artelt2005cross}
\bibfield{author}{\bibinfo{person}{Cordula Artelt}.}
  \bibinfo{year}{2005}\natexlab{}.
\newblock \showarticletitle{Cross-cultural approaches to measuring motivation}.
\newblock \bibinfo{journal}{\emph{Educational Assessment}}
  \bibinfo{volume}{10}, \bibinfo{number}{3} (\bibinfo{year}{2005}),
  \bibinfo{pages}{231--255}.
\newblock


\bibitem[\protect\citeauthoryear{Baker}{Baker}{2012}]%
        {baker2012optimal}
\bibfield{author}{\bibinfo{person}{Elise Baker}.}
  \bibinfo{year}{2012}\natexlab{}.
\newblock \showarticletitle{Optimal intervention intensity}.
\newblock \bibinfo{journal}{\emph{International Journal of Speech-Language
  Pathology}} \bibinfo{volume}{14}, \bibinfo{number}{5} (\bibinfo{year}{2012}),
  \bibinfo{pages}{401--409}.
\newblock


\bibitem[\protect\citeauthoryear{Bandura}{Bandura}{1977}]%
        {bandura1977self}
\bibfield{author}{\bibinfo{person}{Albert Bandura}.}
  \bibinfo{year}{1977}\natexlab{}.
\newblock \showarticletitle{Self-efficacy: toward a unifying theory of
  behavioral change.}
\newblock \bibinfo{journal}{\emph{Psychological review}} \bibinfo{volume}{84},
  \bibinfo{number}{2} (\bibinfo{year}{1977}), \bibinfo{pages}{191}.
\newblock


\bibitem[\protect\citeauthoryear{Barata, Castro, and
  Martins-Lou{\c{c}}{\~a}o}{Barata et~al\mbox{.}}{2017}]%
        {barata2017promote}
\bibfield{author}{\bibinfo{person}{Raquel Barata}, \bibinfo{person}{Paula
  Castro}, {and} \bibinfo{person}{Maria~Am{\'e}lia Martins-Lou{\c{c}}{\~a}o}.}
  \bibinfo{year}{2017}\natexlab{}.
\newblock \showarticletitle{How to promote conservation behaviours: the
  combined role of environmental education and commitment}.
\newblock \bibinfo{journal}{\emph{Environmental Education Research}}
  \bibinfo{volume}{23}, \bibinfo{number}{9} (\bibinfo{year}{2017}),
  \bibinfo{pages}{1322--1334}.
\newblock


\bibitem[\protect\citeauthoryear{Barrett and Barrett}{Barrett and
  Barrett}{2001}]%
        {barrett2001introduction}
\bibfield{author}{\bibinfo{person}{Lisa~Feldman Barrett} {and}
  \bibinfo{person}{Daniel~J Barrett}.} \bibinfo{year}{2001}\natexlab{}.
\newblock \showarticletitle{An introduction to computerized experience sampling
  in psychology}.
\newblock \bibinfo{journal}{\emph{Social Science Computer Review}}
  \bibinfo{volume}{19}, \bibinfo{number}{2} (\bibinfo{year}{2001}),
  \bibinfo{pages}{175--185}.
\newblock


\bibitem[\protect\citeauthoryear{Baumer, Adams, Khovanskaya, Liao, Smith,
  Schwanda~Sosik, and Williams}{Baumer et~al\mbox{.}}{2013}]%
        {baumer2013limiting}
\bibfield{author}{\bibinfo{person}{Eric~PS Baumer}, \bibinfo{person}{Phil
  Adams}, \bibinfo{person}{Vera~D Khovanskaya}, \bibinfo{person}{Tony~C Liao},
  \bibinfo{person}{Madeline~E Smith}, \bibinfo{person}{Victoria
  Schwanda~Sosik}, {and} \bibinfo{person}{Kaiton Williams}.}
  \bibinfo{year}{2013}\natexlab{}.
\newblock \showarticletitle{Limiting, leaving, and (re) lapsing: an exploration
  of facebook non-use practices and experiences}. In
  \bibinfo{booktitle}{\emph{Proceedings of the SIGCHI conference on human
  factors in computing systems}}. \bibinfo{pages}{3257--3266}.
\newblock


\bibitem[\protect\citeauthoryear{Belita and Sidani}{Belita and Sidani}{2015}]%
        {belita2015attrition}
\bibfield{author}{\bibinfo{person}{Emily Belita} {and} \bibinfo{person}{Souraya
  Sidani}.} \bibinfo{year}{2015}\natexlab{}.
\newblock \showarticletitle{Attrition in smoking cessation intervention
  studies: a systematic review}.
\newblock \bibinfo{journal}{\emph{Canadian Journal of Nursing Research
  Archive}} \bibinfo{volume}{47}, \bibinfo{number}{4} (\bibinfo{year}{2015}).
\newblock


\bibitem[\protect\citeauthoryear{Bouton}{Bouton}{2000}]%
        {bouton2000learning}
\bibfield{author}{\bibinfo{person}{Mark~E Bouton}.}
  \bibinfo{year}{2000}\natexlab{}.
\newblock \showarticletitle{A learning theory perspective on lapse, relapse,
  and the maintenance of behavior change.}
\newblock \bibinfo{journal}{\emph{Health Psychology}} \bibinfo{volume}{19},
  \bibinfo{number}{1S} (\bibinfo{year}{2000}), \bibinfo{pages}{57}.
\newblock


\bibitem[\protect\citeauthoryear{Bouton}{Bouton}{2014}]%
        {bouton2014behavior}
\bibfield{author}{\bibinfo{person}{Mark~E Bouton}.}
  \bibinfo{year}{2014}\natexlab{}.
\newblock \showarticletitle{Why behavior change is difficult to sustain}.
\newblock \bibinfo{journal}{\emph{Preventive medicine}}  \bibinfo{volume}{68}
  (\bibinfo{year}{2014}), \bibinfo{pages}{29--36}.
\newblock


\bibitem[\protect\citeauthoryear{Bryan, Karlan, and Nelson}{Bryan
  et~al\mbox{.}}{2010}]%
        {bryan2010commitment}
\bibfield{author}{\bibinfo{person}{Gharad Bryan}, \bibinfo{person}{Dean
  Karlan}, {and} \bibinfo{person}{Scott Nelson}.}
  \bibinfo{year}{2010}\natexlab{}.
\newblock \showarticletitle{Commitment devices}.
\newblock \bibinfo{journal}{\emph{Annu. Rev. Econ.}} \bibinfo{volume}{2},
  \bibinfo{number}{1} (\bibinfo{year}{2010}), \bibinfo{pages}{671--698}.
\newblock


\bibitem[\protect\citeauthoryear{Burgess, Hassm{\'e}n, and Pumpa}{Burgess
  et~al\mbox{.}}{2017}]%
        {burgess2017determinants}
\bibfield{author}{\bibinfo{person}{Emily Burgess}, \bibinfo{person}{Peter
  Hassm{\'e}n}, {and} \bibinfo{person}{Kate~L Pumpa}.}
  \bibinfo{year}{2017}\natexlab{}.
\newblock \showarticletitle{Determinants of adherence to lifestyle intervention
  in adults with obesity: a systematic review}.
\newblock \bibinfo{journal}{\emph{Clinical obesity}} \bibinfo{volume}{7},
  \bibinfo{number}{3} (\bibinfo{year}{2017}), \bibinfo{pages}{123--135}.
\newblock


\bibitem[\protect\citeauthoryear{Burke, Cheng, and de~Gant}{Burke
  et~al\mbox{.}}{2020}]%
        {burke2020social}
\bibfield{author}{\bibinfo{person}{Moira Burke}, \bibinfo{person}{Justin
  Cheng}, {and} \bibinfo{person}{Bethany de Gant}.}
  \bibinfo{year}{2020}\natexlab{}.
\newblock \showarticletitle{Social Comparison and Facebook: Feedback,
  Positivity, and Opportunities for Comparison}. In
  \bibinfo{booktitle}{\emph{Proceedings of the 2020 CHI Conference on Human
  Factors in Computing Systems}}. \bibinfo{pages}{1--13}.
\newblock


\bibitem[\protect\citeauthoryear{Burke and Kraut}{Burke and Kraut}{2016}]%
        {burke2016relationship}
\bibfield{author}{\bibinfo{person}{Moira Burke} {and} \bibinfo{person}{Robert~E
  Kraut}.} \bibinfo{year}{2016}\natexlab{}.
\newblock \showarticletitle{The relationship between Facebook use and
  well-being depends on communication type and tie strength}.
\newblock \bibinfo{journal}{\emph{Journal of computer-mediated communication}}
  \bibinfo{volume}{21}, \bibinfo{number}{4} (\bibinfo{year}{2016}),
  \bibinfo{pages}{265--281}.
\newblock


\bibitem[\protect\citeauthoryear{Cai, Guo, Glass, and Miller}{Cai
  et~al\mbox{.}}{2015}]%
        {cai2015wait}
\bibfield{author}{\bibinfo{person}{Carrie~J Cai}, \bibinfo{person}{Philip~J
  Guo}, \bibinfo{person}{James~R Glass}, {and} \bibinfo{person}{Robert~C
  Miller}.} \bibinfo{year}{2015}\natexlab{}.
\newblock \showarticletitle{Wait-learning: Leveraging wait time for second
  language education}. In \bibinfo{booktitle}{\emph{Proceedings of the 33rd
  Annual ACM Conference on Human Factors in Computing Systems}}.
  \bibinfo{pages}{3701--3710}.
\newblock


\bibitem[\protect\citeauthoryear{Cheng, Burke, and Davis}{Cheng
  et~al\mbox{.}}{2019}]%
        {cheng2019understanding}
\bibfield{author}{\bibinfo{person}{Justin Cheng}, \bibinfo{person}{Moira
  Burke}, {and} \bibinfo{person}{Elena~Goetz Davis}.}
  \bibinfo{year}{2019}\natexlab{}.
\newblock \showarticletitle{Understanding perceptions of problematic Facebook
  use: When people experience negative life impact and a lack of control}. In
  \bibinfo{booktitle}{\emph{Proceedings of the 2019 CHI Conference on Human
  Factors in Computing Systems}}. \bibinfo{pages}{1--13}.
\newblock


\bibitem[\protect\citeauthoryear{Chou}{Chou}{2019}]%
        {chou2019actionable}
\bibfield{author}{\bibinfo{person}{Yu-kai Chou}.}
  \bibinfo{year}{2019}\natexlab{}.
\newblock \bibinfo{booktitle}{\emph{Actionable gamification: Beyond points,
  badges, and leaderboards}}.
\newblock \bibinfo{publisher}{Packt Publishing Ltd}.
\newblock


\bibitem[\protect\citeauthoryear{Collins, Murphy, and Bierman}{Collins
  et~al\mbox{.}}{2004}]%
        {collins2004conceptual}
\bibfield{author}{\bibinfo{person}{Linda~M Collins}, \bibinfo{person}{Susan~A
  Murphy}, {and} \bibinfo{person}{Karen~L Bierman}.}
  \bibinfo{year}{2004}\natexlab{}.
\newblock \showarticletitle{A conceptual framework for adaptive preventive
  interventions}.
\newblock \bibinfo{journal}{\emph{Prevention science}} \bibinfo{volume}{5},
  \bibinfo{number}{3} (\bibinfo{year}{2004}), \bibinfo{pages}{185--196}.
\newblock


\bibitem[\protect\citeauthoryear{Consolvo, McDonald, Toscos, Chen, Froehlich,
  Harrison, Klasnja, LaMarca, LeGrand, Libby, et~al\mbox{.}}{Consolvo
  et~al\mbox{.}}{2008}]%
        {consolvo2008activity}
\bibfield{author}{\bibinfo{person}{Sunny Consolvo}, \bibinfo{person}{David~W
  McDonald}, \bibinfo{person}{Tammy Toscos}, \bibinfo{person}{Mike~Y Chen},
  \bibinfo{person}{Jon Froehlich}, \bibinfo{person}{Beverly Harrison},
  \bibinfo{person}{Predrag Klasnja}, \bibinfo{person}{Anthony LaMarca},
  \bibinfo{person}{Louis LeGrand}, \bibinfo{person}{Ryan Libby},
  {et~al\mbox{.}}} \bibinfo{year}{2008}\natexlab{}.
\newblock \showarticletitle{Activity sensing in the wild: a field trial of
  ubifit garden}. In \bibinfo{booktitle}{\emph{Proceedings of the SIGCHI
  conference on human factors in computing systems}}.
  \bibinfo{pages}{1797--1806}.
\newblock


\bibitem[\protect\citeauthoryear{Cox}{Cox}{1972}]%
        {cox1972regression}
\bibfield{author}{\bibinfo{person}{David~R Cox}.}
  \bibinfo{year}{1972}\natexlab{}.
\newblock \showarticletitle{Regression models and life-tables}.
\newblock \bibinfo{journal}{\emph{Journal of the Royal Statistical Society:
  Series B (Methodological)}} \bibinfo{volume}{34}, \bibinfo{number}{2}
  (\bibinfo{year}{1972}), \bibinfo{pages}{187--202}.
\newblock


\bibitem[\protect\citeauthoryear{Cronqvist and Thaler}{Cronqvist and
  Thaler}{2004}]%
        {cronqvist2004design}
\bibfield{author}{\bibinfo{person}{Henrik Cronqvist} {and}
  \bibinfo{person}{Richard~H Thaler}.} \bibinfo{year}{2004}\natexlab{}.
\newblock \showarticletitle{Design choices in privatized social-security
  systems: Learning from the Swedish experience}.
\newblock \bibinfo{journal}{\emph{American Economic Review}}
  \bibinfo{volume}{94}, \bibinfo{number}{2} (\bibinfo{year}{2004}),
  \bibinfo{pages}{424--428}.
\newblock


\bibitem[\protect\citeauthoryear{Csikszentmihalyi and Larson}{Csikszentmihalyi
  and Larson}{2014}]%
        {csikszentmihalyi2014validity}
\bibfield{author}{\bibinfo{person}{Mihaly Csikszentmihalyi} {and}
  \bibinfo{person}{Reed Larson}.} \bibinfo{year}{2014}\natexlab{}.
\newblock \showarticletitle{Validity and reliability of the experience-sampling
  method}.
\newblock In \bibinfo{booktitle}{\emph{Flow and the foundations of positive
  psychology}}. \bibinfo{publisher}{Springer}, \bibinfo{pages}{35--54}.
\newblock


\bibitem[\protect\citeauthoryear{Cugelman}{Cugelman}{2013}]%
        {cugelman2013gamification}
\bibfield{author}{\bibinfo{person}{Brian Cugelman}.}
  \bibinfo{year}{2013}\natexlab{}.
\newblock \showarticletitle{Gamification: what it is and why it matters to
  digital health behavior change developers}.
\newblock \bibinfo{journal}{\emph{JMIR serious games}} \bibinfo{volume}{1},
  \bibinfo{number}{1} (\bibinfo{year}{2013}), \bibinfo{pages}{e3}.
\newblock


\bibitem[\protect\citeauthoryear{Czaja}{Czaja}{2020}]%
        {czaja2020setting}
\bibfield{author}{\bibinfo{person}{Sara~J Czaja}.}
  \bibinfo{year}{2020}\natexlab{}.
\newblock \showarticletitle{Setting the Stage: Workplace and Demographic
  Trends}.
\newblock In \bibinfo{booktitle}{\emph{Current and Emerging Trends in Aging and
  Work}}. \bibinfo{publisher}{Springer}, \bibinfo{pages}{3--11}.
\newblock


\bibitem[\protect\citeauthoryear{Dabbish, Mark, and Gonz{\'a}lez}{Dabbish
  et~al\mbox{.}}{2011}]%
        {dabbish2011keep}
\bibfield{author}{\bibinfo{person}{Laura Dabbish}, \bibinfo{person}{Gloria
  Mark}, {and} \bibinfo{person}{V{\'\i}ctor~M Gonz{\'a}lez}.}
  \bibinfo{year}{2011}\natexlab{}.
\newblock \showarticletitle{Why do i keep interrupting myself?: environment,
  habit and self-interruption}. In \bibinfo{booktitle}{\emph{Proceedings of the
  SIGCHI Conference on Human Factors in Computing Systems}}. ACM,
  \bibinfo{pages}{3127--3130}.
\newblock


\bibitem[\protect\citeauthoryear{Davidson and Prkachin}{Davidson and
  Prkachin}{1997}]%
        {davidson1997optimism}
\bibfield{author}{\bibinfo{person}{Karina Davidson} {and}
  \bibinfo{person}{Kenneth Prkachin}.} \bibinfo{year}{1997}\natexlab{}.
\newblock \showarticletitle{Optimism and unrealistic optimism have an
  interacting impact on health-promoting behavior and knowledge changes}.
\newblock \bibinfo{journal}{\emph{Personality and social psychology bulletin}}
  \bibinfo{volume}{23}, \bibinfo{number}{6} (\bibinfo{year}{1997}),
  \bibinfo{pages}{617--625}.
\newblock


\bibitem[\protect\citeauthoryear{De~Lara, Tacoronte, and Ding}{De~Lara
  et~al\mbox{.}}{2006}]%
        {de2006current}
\bibfield{author}{\bibinfo{person}{Pablo Zoghbi~Manrique De~Lara},
  \bibinfo{person}{Domingo~Verano Tacoronte}, {and}
  \bibinfo{person}{Jyh-Ming~Ting Ding}.} \bibinfo{year}{2006}\natexlab{}.
\newblock \showarticletitle{Do current anti-cyberloafing disciplinary practices
  have a replica in research findings?}
\newblock \bibinfo{journal}{\emph{Internet Research}} (\bibinfo{year}{2006}).
\newblock


\bibitem[\protect\citeauthoryear{Duckworth, White, Matteucci, Shearer, and
  Gross}{Duckworth et~al\mbox{.}}{2016}]%
        {duckworth2016stitch}
\bibfield{author}{\bibinfo{person}{Angela~L Duckworth},
  \bibinfo{person}{Rachel~E White}, \bibinfo{person}{Alyssa~J Matteucci},
  \bibinfo{person}{Annie Shearer}, {and} \bibinfo{person}{James~J Gross}.}
  \bibinfo{year}{2016}\natexlab{}.
\newblock \showarticletitle{A stitch in time: Strategic self-control in high
  school and college students.}
\newblock \bibinfo{journal}{\emph{Journal of educational psychology}}
  \bibinfo{volume}{108}, \bibinfo{number}{3} (\bibinfo{year}{2016}),
  \bibinfo{pages}{329}.
\newblock


\bibitem[\protect\citeauthoryear{Ernala, Burke, Leavitt, and Ellison}{Ernala
  et~al\mbox{.}}{2020}]%
        {ernala2020well}
\bibfield{author}{\bibinfo{person}{Sindhu~Kiranmai Ernala},
  \bibinfo{person}{Moira Burke}, \bibinfo{person}{Alex Leavitt}, {and}
  \bibinfo{person}{Nicole~B Ellison}.} \bibinfo{year}{2020}\natexlab{}.
\newblock \showarticletitle{How well do people report time spent on Facebook?
  An evaluation of established survey questions with recommendations}. In
  \bibinfo{booktitle}{\emph{Proceedings of the 2020 CHI Conference on Human
  Factors in Computing Systems}}. \bibinfo{pages}{1--14}.
\newblock


\bibitem[\protect\citeauthoryear{Eysenbach}{Eysenbach}{2005}]%
        {eysenbach2005law}
\bibfield{author}{\bibinfo{person}{Gunther Eysenbach}.}
  \bibinfo{year}{2005}\natexlab{}.
\newblock \showarticletitle{The law of attrition}.
\newblock \bibinfo{journal}{\emph{Journal of medical Internet research}}
  \bibinfo{volume}{7}, \bibinfo{number}{1} (\bibinfo{year}{2005}),
  \bibinfo{pages}{e11}.
\newblock


\bibitem[\protect\citeauthoryear{Festinger}{Festinger}{1962}]%
        {festinger1962theory}
\bibfield{author}{\bibinfo{person}{Leon Festinger}.}
  \bibinfo{year}{1962}\natexlab{}.
\newblock \bibinfo{booktitle}{\emph{A theory of cognitive dissonance}}.
  Vol.~\bibinfo{volume}{2}.
\newblock \bibinfo{publisher}{Stanford university press}.
\newblock


\bibitem[\protect\citeauthoryear{Fife and Fife}{Fife and Fife}{2017}]%
        {fife2017package}
\bibfield{author}{\bibinfo{person}{Dustin Fife} {and}
  \bibinfo{person}{Maintainer~Dustin Fife}.} \bibinfo{year}{2017}\natexlab{}.
\newblock \showarticletitle{Package ‘fifer’}.
\newblock \bibinfo{journal}{\emph{A biostatisticians toolbox for various
  activities, including plotting, data cleanup, and data analysis}}
  (\bibinfo{year}{2017}).
\newblock


\bibitem[\protect\citeauthoryear{Fogg}{Fogg}{2009}]%
        {fogg2009behavior}
\bibfield{author}{\bibinfo{person}{Brian~J Fogg}.}
  \bibinfo{year}{2009}\natexlab{}.
\newblock \showarticletitle{A behavior model for persuasive design}. In
  \bibinfo{booktitle}{\emph{Proceedings of the 4th international Conference on
  Persuasive Technology}}. \bibinfo{pages}{1--7}.
\newblock


\bibitem[\protect\citeauthoryear{Fox, Ratner, and Lieb}{Fox
  et~al\mbox{.}}{2005}]%
        {fox2005subjective}
\bibfield{author}{\bibinfo{person}{Craig~R Fox}, \bibinfo{person}{Rebecca~K
  Ratner}, {and} \bibinfo{person}{Daniel~S Lieb}.}
  \bibinfo{year}{2005}\natexlab{}.
\newblock \showarticletitle{How subjective grouping of options influences
  choice and allocation: diversification bias and the phenomenon of partition
  dependence.}
\newblock \bibinfo{journal}{\emph{Journal of Experimental Psychology: General}}
  \bibinfo{volume}{134}, \bibinfo{number}{4} (\bibinfo{year}{2005}),
  \bibinfo{pages}{538}.
\newblock


\bibitem[\protect\citeauthoryear{Fulmer and Frijters}{Fulmer and
  Frijters}{2009}]%
        {fulmer2009review}
\bibfield{author}{\bibinfo{person}{Sara~M Fulmer} {and} \bibinfo{person}{Jan~C
  Frijters}.} \bibinfo{year}{2009}\natexlab{}.
\newblock \showarticletitle{A review of self-report and alternative approaches
  in the measurement of student motivation}.
\newblock \bibinfo{journal}{\emph{Educational Psychology Review}}
  \bibinfo{volume}{21}, \bibinfo{number}{3} (\bibinfo{year}{2009}),
  \bibinfo{pages}{219--246}.
\newblock


\bibitem[\protect\citeauthoryear{Gin{\'e}, Karlan, and Zinman}{Gin{\'e}
  et~al\mbox{.}}{2010}]%
        {gine2010put}
\bibfield{author}{\bibinfo{person}{Xavier Gin{\'e}}, \bibinfo{person}{Dean
  Karlan}, {and} \bibinfo{person}{Jonathan Zinman}.}
  \bibinfo{year}{2010}\natexlab{}.
\newblock \showarticletitle{Put your money where your butt is: a commitment
  contract for smoking cessation}.
\newblock \bibinfo{journal}{\emph{American Economic Journal: Applied
  Economics}} \bibinfo{volume}{2}, \bibinfo{number}{4} (\bibinfo{year}{2010}),
  \bibinfo{pages}{213--35}.
\newblock


\bibitem[\protect\citeauthoryear{Given, Given, and Coyle}{Given
  et~al\mbox{.}}{1985}]%
        {given1985prediction}
\bibfield{author}{\bibinfo{person}{Charles~W Given}, \bibinfo{person}{Barbara~A
  Given}, {and} \bibinfo{person}{Bryan~W Coyle}.}
  \bibinfo{year}{1985}\natexlab{}.
\newblock \showarticletitle{Prediction of patient attrition from experimental
  behavioral interventions.}
\newblock \bibinfo{journal}{\emph{Nursing research}} (\bibinfo{year}{1985}).
\newblock


\bibitem[\protect\citeauthoryear{Glassman, Prosch, and Shao}{Glassman
  et~al\mbox{.}}{2015}]%
        {glassman2015monitor}
\bibfield{author}{\bibinfo{person}{Jeremy Glassman}, \bibinfo{person}{Marilyn
  Prosch}, {and} \bibinfo{person}{Benjamin~BM Shao}.}
  \bibinfo{year}{2015}\natexlab{}.
\newblock \showarticletitle{To monitor or not to monitor: Effectiveness of a
  cyberloafing countermeasure}.
\newblock \bibinfo{journal}{\emph{Information \& Management}}
  \bibinfo{volume}{52}, \bibinfo{number}{2} (\bibinfo{year}{2015}),
  \bibinfo{pages}{170--182}.
\newblock


\bibitem[\protect\citeauthoryear{Gugerty}{Gugerty}{2007}]%
        {gugerty2007you}
\bibfield{author}{\bibinfo{person}{Mary~Kay Gugerty}.}
  \bibinfo{year}{2007}\natexlab{}.
\newblock \showarticletitle{You can’t save alone: Commitment in rotating
  savings and credit associations in Kenya}.
\newblock \bibinfo{journal}{\emph{Economic Development and cultural change}}
  \bibinfo{volume}{55}, \bibinfo{number}{2} (\bibinfo{year}{2007}),
  \bibinfo{pages}{251--282}.
\newblock


\bibitem[\protect\citeauthoryear{Halpern, Asch, and Volpp}{Halpern
  et~al\mbox{.}}{2012}]%
        {halpern2012commitment}
\bibfield{author}{\bibinfo{person}{Scott~D Halpern}, \bibinfo{person}{David~A
  Asch}, {and} \bibinfo{person}{Kevin~G Volpp}.}
  \bibinfo{year}{2012}\natexlab{}.
\newblock \showarticletitle{Commitment contracts as a way to health}.
\newblock \bibinfo{journal}{\emph{Bmj}}  \bibinfo{volume}{344}
  (\bibinfo{year}{2012}), \bibinfo{pages}{e522}.
\newblock


\bibitem[\protect\citeauthoryear{Hektner, Schmidt, and
  Csikszentmihalyi}{Hektner et~al\mbox{.}}{2007}]%
        {hektner2007experience}
\bibfield{author}{\bibinfo{person}{Joel~M Hektner}, \bibinfo{person}{Jennifer~A
  Schmidt}, {and} \bibinfo{person}{Mihaly Csikszentmihalyi}.}
  \bibinfo{year}{2007}\natexlab{}.
\newblock \bibinfo{booktitle}{\emph{Experience sampling method: Measuring the
  quality of everyday life}}.
\newblock \bibinfo{publisher}{Sage}.
\newblock


\bibitem[\protect\citeauthoryear{Hofmann, Baumeister, F{\"o}rster, and
  Vohs}{Hofmann et~al\mbox{.}}{2012}]%
        {hofmann2012everyday}
\bibfield{author}{\bibinfo{person}{Wilhelm Hofmann}, \bibinfo{person}{Roy~F
  Baumeister}, \bibinfo{person}{Georg F{\"o}rster}, {and}
  \bibinfo{person}{Kathleen~D Vohs}.} \bibinfo{year}{2012}\natexlab{}.
\newblock \showarticletitle{Everyday temptations: an experience sampling study
  of desire, conflict, and self-control.}
\newblock \bibinfo{journal}{\emph{Journal of personality and social
  psychology}} \bibinfo{volume}{102}, \bibinfo{number}{6}
  (\bibinfo{year}{2012}), \bibinfo{pages}{1318}.
\newblock


\bibitem[\protect\citeauthoryear{Huynh and Iida}{Huynh and Iida}{2017}]%
        {huynh2017analysis}
\bibfield{author}{\bibinfo{person}{Duy Huynh} {and} \bibinfo{person}{Hiroyuki
  Iida}.} \bibinfo{year}{2017}\natexlab{}.
\newblock \showarticletitle{An analysis of winning streak's effects in language
  course of “Duolingo”}.
\newblock \bibinfo{journal}{\emph{Asia-Pacific Journal of Information
  Technology and Multimedia}} \bibinfo{volume}{6}, \bibinfo{number}{2}
  (\bibinfo{year}{2017}).
\newblock


\bibitem[\protect\citeauthoryear{Iuga and McGuire}{Iuga and McGuire}{2014}]%
        {iuga2014adherence}
\bibfield{author}{\bibinfo{person}{Aurel~O Iuga} {and} \bibinfo{person}{Maura~J
  McGuire}.} \bibinfo{year}{2014}\natexlab{}.
\newblock \showarticletitle{Adherence and health care costs}.
\newblock \bibinfo{journal}{\emph{Risk management and healthcare policy}}
  \bibinfo{volume}{7} (\bibinfo{year}{2014}), \bibinfo{pages}{35}.
\newblock


\bibitem[\protect\citeauthoryear{Jin and Dabbish}{Jin and Dabbish}{2009}]%
        {jin2009self}
\bibfield{author}{\bibinfo{person}{Jing Jin} {and} \bibinfo{person}{Laura~A
  Dabbish}.} \bibinfo{year}{2009}\natexlab{}.
\newblock \showarticletitle{Self-interruption on the computer: a typology of
  discretionary task interleaving}. In \bibinfo{booktitle}{\emph{Proceedings of
  the SIGCHI conference on human factors in computing systems}}. ACM,
  \bibinfo{pages}{1799--1808}.
\newblock


\bibitem[\protect\citeauthoryear{Johnson, Shu, Dellaert, Fox, Goldstein,
  H{\"a}ubl, Larrick, Payne, Peters, Schkade, et~al\mbox{.}}{Johnson
  et~al\mbox{.}}{2012}]%
        {johnson2012beyond}
\bibfield{author}{\bibinfo{person}{Eric~J Johnson}, \bibinfo{person}{Suzanne~B
  Shu}, \bibinfo{person}{Benedict~GC Dellaert}, \bibinfo{person}{Craig Fox},
  \bibinfo{person}{Daniel~G Goldstein}, \bibinfo{person}{Gerald H{\"a}ubl},
  \bibinfo{person}{Richard~P Larrick}, \bibinfo{person}{John~W Payne},
  \bibinfo{person}{Ellen Peters}, \bibinfo{person}{David Schkade},
  {et~al\mbox{.}}} \bibinfo{year}{2012}\natexlab{}.
\newblock \showarticletitle{Beyond nudges: Tools of a choice architecture}.
\newblock \bibinfo{journal}{\emph{Marketing Letters}} \bibinfo{volume}{23},
  \bibinfo{number}{2} (\bibinfo{year}{2012}), \bibinfo{pages}{487--504}.
\newblock


\bibitem[\protect\citeauthoryear{Junco}{Junco}{2013}]%
        {junco2013comparing}
\bibfield{author}{\bibinfo{person}{Reynol Junco}.}
  \bibinfo{year}{2013}\natexlab{}.
\newblock \showarticletitle{Comparing actual and self-reported measures of
  Facebook use}.
\newblock \bibinfo{journal}{\emph{Computers in Human Behavior}}
  \bibinfo{volume}{29}, \bibinfo{number}{3} (\bibinfo{year}{2013}),
  \bibinfo{pages}{626--631}.
\newblock


\bibitem[\protect\citeauthoryear{Kahneman and Lovallo}{Kahneman and
  Lovallo}{1993}]%
        {kahneman1993timid}
\bibfield{author}{\bibinfo{person}{Daniel Kahneman} {and} \bibinfo{person}{Dan
  Lovallo}.} \bibinfo{year}{1993}\natexlab{}.
\newblock \showarticletitle{Timid choices and bold forecasts: A cognitive
  perspective on risk taking}.
\newblock \bibinfo{journal}{\emph{Management science}} \bibinfo{volume}{39},
  \bibinfo{number}{1} (\bibinfo{year}{1993}), \bibinfo{pages}{17--31}.
\newblock


\bibitem[\protect\citeauthoryear{Kim, Cho, and Lee}{Kim et~al\mbox{.}}{2017a}]%
        {kim2017technology}
\bibfield{author}{\bibinfo{person}{Jaejeung Kim}, \bibinfo{person}{Chiwoo Cho},
  {and} \bibinfo{person}{Uichin Lee}.} \bibinfo{year}{2017}\natexlab{a}.
\newblock \showarticletitle{Technology supported behavior restriction for
  mitigating self-interruptions in multi-device environments}.
\newblock \bibinfo{journal}{\emph{Proceedings of the ACM on Interactive,
  Mobile, Wearable and Ubiquitous Technologies}} \bibinfo{volume}{1},
  \bibinfo{number}{3} (\bibinfo{year}{2017}), \bibinfo{pages}{1--21}.
\newblock


\bibitem[\protect\citeauthoryear{Kim, Jeon, Choe, Lee, Kim, and Seo}{Kim
  et~al\mbox{.}}{2016}]%
        {kim2016timeaware}
\bibfield{author}{\bibinfo{person}{Young-Ho Kim}, \bibinfo{person}{Jae~Ho
  Jeon}, \bibinfo{person}{Eun~Kyoung Choe}, \bibinfo{person}{Bongshin Lee},
  \bibinfo{person}{KwonHyun Kim}, {and} \bibinfo{person}{Jinwook Seo}.}
  \bibinfo{year}{2016}\natexlab{}.
\newblock \showarticletitle{TimeAware: Leveraging framing effects to enhance
  personal productivity}. In \bibinfo{booktitle}{\emph{Proceedings of the 2016
  CHI Conference on Human Factors in Computing Systems}}.
  \bibinfo{pages}{272--283}.
\newblock


\bibitem[\protect\citeauthoryear{Kim, Jeon, Lee, Choe, and Seo}{Kim
  et~al\mbox{.}}{2017b}]%
        {kim2017omnitrack}
\bibfield{author}{\bibinfo{person}{Young-Ho Kim}, \bibinfo{person}{Jae~Ho
  Jeon}, \bibinfo{person}{Bongshin Lee}, \bibinfo{person}{Eun~Kyoung Choe},
  {and} \bibinfo{person}{Jinwook Seo}.} \bibinfo{year}{2017}\natexlab{b}.
\newblock \showarticletitle{OmniTrack: A flexible self-tracking approach
  leveraging semi-automated tracking}.
\newblock \bibinfo{journal}{\emph{Proceedings of the ACM on Interactive,
  Mobile, Wearable and Ubiquitous Technologies}} \bibinfo{volume}{1},
  \bibinfo{number}{3} (\bibinfo{year}{2017}), \bibinfo{pages}{1--28}.
\newblock


\bibitem[\protect\citeauthoryear{Koehler}{Koehler}{1991}]%
        {koehler1991explanation}
\bibfield{author}{\bibinfo{person}{Derek~J Koehler}.}
  \bibinfo{year}{1991}\natexlab{}.
\newblock \showarticletitle{Explanation, imagination, and confidence in
  judgment.}
\newblock \bibinfo{journal}{\emph{Psychological bulletin}}
  \bibinfo{volume}{110}, \bibinfo{number}{3} (\bibinfo{year}{1991}),
  \bibinfo{pages}{499}.
\newblock


\bibitem[\protect\citeauthoryear{Kovacs}{Kovacs}{2015}]%
        {kovacs2015feedlearn}
\bibfield{author}{\bibinfo{person}{Geza Kovacs}.}
  \bibinfo{year}{2015}\natexlab{}.
\newblock \showarticletitle{FeedLearn: Using facebook feeds for microlearning}.
  In \bibinfo{booktitle}{\emph{Proceedings of the 33rd annual ACM Conference
  extended abstracts on human factors in computing systems}}.
  \bibinfo{pages}{1461--1466}.
\newblock


\bibitem[\protect\citeauthoryear{Kovacs, Gregory, Ma, Wu, Emami, Ray, and
  Bernstein}{Kovacs et~al\mbox{.}}{2019}]%
        {kovacs2019conservation}
\bibfield{author}{\bibinfo{person}{Geza Kovacs}, \bibinfo{person}{Drew~Mylander
  Gregory}, \bibinfo{person}{Zilin Ma}, \bibinfo{person}{Zhengxuan Wu},
  \bibinfo{person}{Golrokh Emami}, \bibinfo{person}{Jacob Ray}, {and}
  \bibinfo{person}{Michael~S Bernstein}.} \bibinfo{year}{2019}\natexlab{}.
\newblock \showarticletitle{Conservation of Procrastination: Do Productivity
  Interventions Save Time or Just Redistribute It?}. In
  \bibinfo{booktitle}{\emph{Proceedings of the 2019 CHI Conference on Human
  Factors in Computing Systems}}. ACM, \bibinfo{pages}{330}.
\newblock


\bibitem[\protect\citeauthoryear{Kovacs, Wu, and Bernstein}{Kovacs
  et~al\mbox{.}}{2018}]%
        {kovacs2018rotating}
\bibfield{author}{\bibinfo{person}{Geza Kovacs}, \bibinfo{person}{Zhengxuan
  Wu}, {and} \bibinfo{person}{Michael~S Bernstein}.}
  \bibinfo{year}{2018}\natexlab{}.
\newblock \showarticletitle{Rotating online behavior change interventions
  increases effectiveness but also increases attrition}.
\newblock \bibinfo{journal}{\emph{Proceedings of the ACM on Human-Computer
  Interaction}} \bibinfo{volume}{2}, \bibinfo{number}{CSCW}
  (\bibinfo{year}{2018}), \bibinfo{pages}{1--25}.
\newblock


\bibitem[\protect\citeauthoryear{Kraushaar and Novak}{Kraushaar and
  Novak}{2019}]%
        {kraushaar2019examining}
\bibfield{author}{\bibinfo{person}{James~M Kraushaar} {and}
  \bibinfo{person}{David~C Novak}.} \bibinfo{year}{2019}\natexlab{}.
\newblock \showarticletitle{Examining the affects of student multitasking with
  laptops during the lecture}.
\newblock \bibinfo{journal}{\emph{Journal of Information Systems Education}}
  \bibinfo{volume}{21}, \bibinfo{number}{2} (\bibinfo{year}{2019}),
  \bibinfo{pages}{11}.
\newblock


\bibitem[\protect\citeauthoryear{Kross, Verduyn, Demiralp, Park, Lee, Lin,
  Shablack, Jonides, and Ybarra}{Kross et~al\mbox{.}}{2013}]%
        {kross2013facebook}
\bibfield{author}{\bibinfo{person}{Ethan Kross}, \bibinfo{person}{Philippe
  Verduyn}, \bibinfo{person}{Emre Demiralp}, \bibinfo{person}{Jiyoung Park},
  \bibinfo{person}{David~Seungjae Lee}, \bibinfo{person}{Natalie Lin},
  \bibinfo{person}{Holly Shablack}, \bibinfo{person}{John Jonides}, {and}
  \bibinfo{person}{Oscar Ybarra}.} \bibinfo{year}{2013}\natexlab{}.
\newblock \showarticletitle{Facebook use predicts declines in subjective
  well-being in young adults}.
\newblock \bibinfo{journal}{\emph{PloS one}} \bibinfo{volume}{8},
  \bibinfo{number}{8} (\bibinfo{year}{2013}), \bibinfo{pages}{e69841}.
\newblock


\bibitem[\protect\citeauthoryear{K{\"u}nzler}{K{\"u}nzler}{2019}]%
        {kunzler2019context}
\bibfield{author}{\bibinfo{person}{Florian K{\"u}nzler}.}
  \bibinfo{year}{2019}\natexlab{}.
\newblock \showarticletitle{Context-aware notification management systems for
  just-in-time adaptive interventions}. In \bibinfo{booktitle}{\emph{2019 IEEE
  International Conference on Pervasive Computing and Communications Workshops
  (PerCom Workshops)}}. IEEE, \bibinfo{pages}{435--436}.
\newblock


\bibitem[\protect\citeauthoryear{Kwee, Komoru-Venovic, and Kwee}{Kwee
  et~al\mbox{.}}{2010}]%
        {kwee2010treatment}
\bibfield{author}{\bibinfo{person}{Alex~W Kwee}, \bibinfo{person}{E
  Komoru-Venovic}, {and} \bibinfo{person}{Janelle~L Kwee}.}
  \bibinfo{year}{2010}\natexlab{}.
\newblock \showarticletitle{Treatment implications from Etiological and
  diagnostic considerations of internet Addiction: cautions with the Boot camp
  Approach}. In \bibinfo{booktitle}{\emph{Proceedings of the International
  Conference of e-CASE, Distinguished Paper, CD Format, Singapore}}.
\newblock


\bibitem[\protect\citeauthoryear{Leotti, Iyengar, and Ochsner}{Leotti
  et~al\mbox{.}}{2010}]%
        {leotti2010born}
\bibfield{author}{\bibinfo{person}{Lauren~A Leotti}, \bibinfo{person}{Sheena~S
  Iyengar}, {and} \bibinfo{person}{Kevin~N Ochsner}.}
  \bibinfo{year}{2010}\natexlab{}.
\newblock \showarticletitle{Born to choose: The origins and value of the need
  for control}.
\newblock \bibinfo{journal}{\emph{Trends in cognitive sciences}}
  \bibinfo{volume}{14}, \bibinfo{number}{10} (\bibinfo{year}{2010}),
  \bibinfo{pages}{457--463}.
\newblock


\bibitem[\protect\citeauthoryear{Lim and Chen}{Lim and Chen}{2012}]%
        {lim2012cyberloafing}
\bibfield{author}{\bibinfo{person}{Vivien~KG Lim} {and} \bibinfo{person}{Don~JQ
  Chen}.} \bibinfo{year}{2012}\natexlab{}.
\newblock \showarticletitle{Cyberloafing at the workplace: gain or drain on
  work?}
\newblock \bibinfo{journal}{\emph{Behaviour \& Information Technology}}
  \bibinfo{volume}{31}, \bibinfo{number}{4} (\bibinfo{year}{2012}),
  \bibinfo{pages}{343--353}.
\newblock


\bibitem[\protect\citeauthoryear{Linardon and Fuller-Tyszkiewicz}{Linardon and
  Fuller-Tyszkiewicz}{2020}]%
        {linardon2020attrition}
\bibfield{author}{\bibinfo{person}{Jake Linardon} {and}
  \bibinfo{person}{Matthew Fuller-Tyszkiewicz}.}
  \bibinfo{year}{2020}\natexlab{}.
\newblock \showarticletitle{Attrition and adherence in smartphone-delivered
  interventions for mental health problems: A systematic and meta-analytic
  review.}
\newblock \bibinfo{journal}{\emph{Journal of consulting and clinical
  psychology}} \bibinfo{volume}{88}, \bibinfo{number}{1}
  (\bibinfo{year}{2020}), \bibinfo{pages}{1}.
\newblock


\bibitem[\protect\citeauthoryear{Lister and Harnish}{Lister and
  Harnish}{2019}]%
        {lister2019telework}
\bibfield{author}{\bibinfo{person}{Kate Lister} {and} \bibinfo{person}{Tom
  Harnish}.} \bibinfo{year}{2019}\natexlab{}.
\newblock \showarticletitle{Telework and its effects in the United States}.
\newblock In \bibinfo{booktitle}{\emph{Telework in the 21st Century}}.
  \bibinfo{publisher}{Edward Elgar Publishing}.
\newblock


\bibitem[\protect\citeauthoryear{Liu, Liu, Ding, and Yang}{Liu
  et~al\mbox{.}}{2013}]%
        {liu2013adherence}
\bibfield{author}{\bibinfo{person}{Jianming Liu}, \bibinfo{person}{Zhiliang
  Liu}, \bibinfo{person}{Hu Ding}, {and} \bibinfo{person}{Xiaohong Yang}.}
  \bibinfo{year}{2013}\natexlab{}.
\newblock \showarticletitle{Adherence to treatment and influencing factors in a
  sample of Chinese epilepsy patients}.
\newblock \bibinfo{journal}{\emph{Epileptic disorders}} \bibinfo{volume}{15},
  \bibinfo{number}{3} (\bibinfo{year}{2013}), \bibinfo{pages}{289--294}.
\newblock


\bibitem[\protect\citeauthoryear{Loewenstein and Elster}{Loewenstein and
  Elster}{1992}]%
        {loewenstein1992choice}
\bibfield{author}{\bibinfo{person}{George Loewenstein} {and}
  \bibinfo{person}{Jon Elster}.} \bibinfo{year}{1992}\natexlab{}.
\newblock \bibinfo{booktitle}{\emph{Choice over time}}.
\newblock \bibinfo{publisher}{Russell Sage Foundation}.
\newblock


\bibitem[\protect\citeauthoryear{Loewenstein and Schkade}{Loewenstein and
  Schkade}{1999}]%
        {loewenstein1999wouldn}
\bibfield{author}{\bibinfo{person}{George Loewenstein} {and}
  \bibinfo{person}{David Schkade}.} \bibinfo{year}{1999}\natexlab{}.
\newblock \showarticletitle{Wouldn’t it be nice? Predicting future feelings}.
\newblock \bibinfo{journal}{\emph{Well-being: The foundations of hedonic
  psychology}} (\bibinfo{year}{1999}), \bibinfo{pages}{85--105}.
\newblock


\bibitem[\protect\citeauthoryear{Luettgau, Tempelmann, Kaiser, and
  Jocham}{Luettgau et~al\mbox{.}}{2020}]%
        {luettgau2020decisions}
\bibfield{author}{\bibinfo{person}{Lennart Luettgau}, \bibinfo{person}{Claus
  Tempelmann}, \bibinfo{person}{Luca~Franziska Kaiser}, {and}
  \bibinfo{person}{Gerhard Jocham}.} \bibinfo{year}{2020}\natexlab{}.
\newblock \showarticletitle{Decisions bias future choices by modifying
  hippocampal associative memories}.
\newblock \bibinfo{journal}{\emph{Nature communications}} \bibinfo{volume}{11},
  \bibinfo{number}{1} (\bibinfo{year}{2020}), \bibinfo{pages}{1--14}.
\newblock


\bibitem[\protect\citeauthoryear{Lynch~Jr and Ariely}{Lynch~Jr and
  Ariely}{2000}]%
        {lynch2000wine}
\bibfield{author}{\bibinfo{person}{John~G Lynch~Jr} {and} \bibinfo{person}{Dan
  Ariely}.} \bibinfo{year}{2000}\natexlab{}.
\newblock \showarticletitle{Wine online: Search costs affect competition on
  price, quality, and distribution}.
\newblock \bibinfo{journal}{\emph{Marketing science}} \bibinfo{volume}{19},
  \bibinfo{number}{1} (\bibinfo{year}{2000}), \bibinfo{pages}{83--103}.
\newblock


\bibitem[\protect\citeauthoryear{Lyngs, Lukoff, Slovak, Binns, Slack, Inzlicht,
  Van~Kleek, and Shadbolt}{Lyngs et~al\mbox{.}}{2019}]%
        {lyngs2019self}
\bibfield{author}{\bibinfo{person}{Ulrik Lyngs}, \bibinfo{person}{Kai Lukoff},
  \bibinfo{person}{Petr Slovak}, \bibinfo{person}{Reuben Binns},
  \bibinfo{person}{Adam Slack}, \bibinfo{person}{Michael Inzlicht},
  \bibinfo{person}{Max Van~Kleek}, {and} \bibinfo{person}{Nigel Shadbolt}.}
  \bibinfo{year}{2019}\natexlab{}.
\newblock \showarticletitle{Self-Control in Cyberspace: Applying Dual Systems
  Theory to a Review of Digital Self-Control Tools}. In
  \bibinfo{booktitle}{\emph{Proceedings of the 2019 CHI Conference on Human
  Factors in Computing Systems}}. \bibinfo{pages}{1--18}.
\newblock


\bibitem[\protect\citeauthoryear{Lyngs, Lukoff, Slovak, Seymour, Webb, Jirotka,
  Zhao, Van~Kleek, and Shadbolt}{Lyngs et~al\mbox{.}}{2020}]%
        {lyngs2020just}
\bibfield{author}{\bibinfo{person}{Ulrik Lyngs}, \bibinfo{person}{Kai Lukoff},
  \bibinfo{person}{Petr Slovak}, \bibinfo{person}{William Seymour},
  \bibinfo{person}{Helena Webb}, \bibinfo{person}{Marina Jirotka},
  \bibinfo{person}{Jun Zhao}, \bibinfo{person}{Max Van~Kleek}, {and}
  \bibinfo{person}{Nigel Shadbolt}.} \bibinfo{year}{2020}\natexlab{}.
\newblock \showarticletitle{'I Just Want to Hack Myself to Not Get Distracted'
  Evaluating Design Interventions for Self-Control on Facebook}. In
  \bibinfo{booktitle}{\emph{Proceedings of the 2020 CHI Conference on Human
  Factors in Computing Systems}}. \bibinfo{pages}{1--15}.
\newblock


\bibitem[\protect\citeauthoryear{Mark, Czerwinski, and Iqbal}{Mark
  et~al\mbox{.}}{2018}]%
        {mark2018effects}
\bibfield{author}{\bibinfo{person}{Gloria Mark}, \bibinfo{person}{Mary
  Czerwinski}, {and} \bibinfo{person}{Shamsi~T Iqbal}.}
  \bibinfo{year}{2018}\natexlab{}.
\newblock \showarticletitle{Effects of individual differences in blocking
  workplace distractions}. In \bibinfo{booktitle}{\emph{Proceedings of the 2018
  CHI Conference on Human Factors in Computing Systems}}.
  \bibinfo{pages}{1--12}.
\newblock


\bibitem[\protect\citeauthoryear{Mark, Iqbal, Czerwinski, and Johns}{Mark
  et~al\mbox{.}}{2015}]%
        {mark2015focused}
\bibfield{author}{\bibinfo{person}{Gloria Mark}, \bibinfo{person}{Shamsi
  Iqbal}, \bibinfo{person}{Mary Czerwinski}, {and} \bibinfo{person}{Paul
  Johns}.} \bibinfo{year}{2015}\natexlab{}.
\newblock \showarticletitle{Focused, aroused, but so distractible: Temporal
  perspectives on multitasking and communications}. In
  \bibinfo{booktitle}{\emph{Proceedings of the 18th ACM Conference on Computer
  Supported Cooperative Work \& Social Computing}}. \bibinfo{pages}{903--916}.
\newblock


\bibitem[\protect\citeauthoryear{Mark, Iqbal, Czerwinski, Johns, Sano, and
  Lutchyn}{Mark et~al\mbox{.}}{2016}]%
        {mark2016email}
\bibfield{author}{\bibinfo{person}{Gloria Mark}, \bibinfo{person}{Shamsi~T
  Iqbal}, \bibinfo{person}{Mary Czerwinski}, \bibinfo{person}{Paul Johns},
  \bibinfo{person}{Akane Sano}, {and} \bibinfo{person}{Yuliya Lutchyn}.}
  \bibinfo{year}{2016}\natexlab{}.
\newblock \showarticletitle{Email duration, batching and self-interruption:
  Patterns of email use on productivity and stress}. In
  \bibinfo{booktitle}{\emph{Proceedings of the 2016 CHI Conference on Human
  Factors in Computing Systems}}. \bibinfo{pages}{1717--1728}.
\newblock


\bibitem[\protect\citeauthoryear{Mark, Wang, and Niiya}{Mark
  et~al\mbox{.}}{2014}]%
        {mark2014stress}
\bibfield{author}{\bibinfo{person}{Gloria Mark}, \bibinfo{person}{Yiran Wang},
  {and} \bibinfo{person}{Melissa Niiya}.} \bibinfo{year}{2014}\natexlab{}.
\newblock \showarticletitle{Stress and multitasking in everyday college life:
  an empirical study of online activity}. In
  \bibinfo{booktitle}{\emph{Proceedings of the SIGCHI Conference on Human
  Factors in Computing Systems}}. \bibinfo{pages}{41--50}.
\newblock


\bibitem[\protect\citeauthoryear{Marlatt}{Marlatt}{1996}]%
        {marlatt1996taxonomy}
\bibfield{author}{\bibinfo{person}{G~Alan Marlatt}.}
  \bibinfo{year}{1996}\natexlab{}.
\newblock \showarticletitle{Taxonomy of high-risk situations for alcohol
  relapse: evolution and development of a}.
\newblock \bibinfo{journal}{\emph{Addiction}} \bibinfo{volume}{91},
  \bibinfo{number}{12s1} (\bibinfo{year}{1996}), \bibinfo{pages}{37--50}.
\newblock


\bibitem[\protect\citeauthoryear{Marlatt, Baer, and Quigley}{Marlatt
  et~al\mbox{.}}{1997}]%
        {marlatt199710}
\bibfield{author}{\bibinfo{person}{G~Alan Marlatt}, \bibinfo{person}{John~S
  Baer}, {and} \bibinfo{person}{Lori~A Quigley}.}
  \bibinfo{year}{1997}\natexlab{}.
\newblock \showarticletitle{Self-efficacy and addictive behavior}.
\newblock \bibinfo{journal}{\emph{Selfefficacy in changing societies}}
  (\bibinfo{year}{1997}), \bibinfo{pages}{289--315}.
\newblock


\bibitem[\protect\citeauthoryear{Martin}{Martin}{2001}]%
        {martin2001student}
\bibfield{author}{\bibinfo{person}{Andrew~J Martin}.}
  \bibinfo{year}{2001}\natexlab{}.
\newblock \showarticletitle{The Student Motivation Scale: A tool for measuring
  and enhancing motivation}.
\newblock \bibinfo{journal}{\emph{Australian Journal of guidance and
  Counselling}} \bibinfo{volume}{11}, \bibinfo{number}{1}
  (\bibinfo{year}{2001}), \bibinfo{pages}{1--20}.
\newblock


\bibitem[\protect\citeauthoryear{Martin, Bassi, and Dunbar-Rees}{Martin
  et~al\mbox{.}}{2012}]%
        {martin2012commitments}
\bibfield{author}{\bibinfo{person}{Steve~J Martin}, \bibinfo{person}{Suraj
  Bassi}, {and} \bibinfo{person}{Rupert Dunbar-Rees}.}
  \bibinfo{year}{2012}\natexlab{}.
\newblock \showarticletitle{Commitments, norms and custard creams--a social
  influence approach to reducing did not attends (DNAs)}.
\newblock \bibinfo{journal}{\emph{Journal of the Royal Society of Medicine}}
  \bibinfo{volume}{105}, \bibinfo{number}{3} (\bibinfo{year}{2012}),
  \bibinfo{pages}{101--104}.
\newblock


\bibitem[\protect\citeauthoryear{Marx}{Marx}{1982}]%
        {marx1982relapse}
\bibfield{author}{\bibinfo{person}{Robert~D Marx}.}
  \bibinfo{year}{1982}\natexlab{}.
\newblock \showarticletitle{Relapse prevention for managerial training: A model
  for maintenance of behavior change}.
\newblock \bibinfo{journal}{\emph{Academy of management review}}
  \bibinfo{volume}{7}, \bibinfo{number}{3} (\bibinfo{year}{1982}),
  \bibinfo{pages}{433--441}.
\newblock


\bibitem[\protect\citeauthoryear{McAuley, Poag, Gleason, and Wraith}{McAuley
  et~al\mbox{.}}{1990}]%
        {mcauley1990attrition}
\bibfield{author}{\bibinfo{person}{Edward McAuley}, \bibinfo{person}{Kim Poag},
  \bibinfo{person}{Anita Gleason}, {and} \bibinfo{person}{Susan Wraith}.}
  \bibinfo{year}{1990}\natexlab{}.
\newblock \showarticletitle{Attrition from exercise programs: Attributional and
  affective perspectives}.
\newblock \bibinfo{journal}{\emph{Journal of Social Behavior and Personality}}
  \bibinfo{volume}{5}, \bibinfo{number}{6} (\bibinfo{year}{1990}),
  \bibinfo{pages}{591}.
\newblock


\bibitem[\protect\citeauthoryear{Meier, Reinecke, and Meltzer}{Meier
  et~al\mbox{.}}{2016}]%
        {meier2016facebocrastination}
\bibfield{author}{\bibinfo{person}{Adrian Meier}, \bibinfo{person}{Leonard
  Reinecke}, {and} \bibinfo{person}{Christine~E Meltzer}.}
  \bibinfo{year}{2016}\natexlab{}.
\newblock \showarticletitle{“Facebocrastination”? Predictors of using
  Facebook for procrastination and its effects on students’ well-being}.
\newblock \bibinfo{journal}{\emph{Computers in Human Behavior}}
  \bibinfo{volume}{64} (\bibinfo{year}{2016}), \bibinfo{pages}{65--76}.
\newblock


\bibitem[\protect\citeauthoryear{Michael, Goldberg, Treuth, Beans, Normandt,
  and Macko}{Michael et~al\mbox{.}}{2009}]%
        {michael2009progressive}
\bibfield{author}{\bibinfo{person}{Kathleen Michael}, \bibinfo{person}{Andrew~P
  Goldberg}, \bibinfo{person}{Margarita~S Treuth}, \bibinfo{person}{Jeffrey
  Beans}, \bibinfo{person}{Peter Normandt}, {and} \bibinfo{person}{Richard~F
  Macko}.} \bibinfo{year}{2009}\natexlab{}.
\newblock \showarticletitle{Progressive adaptive physical activity in stroke
  improves balance, gait, and fitness: preliminary results}.
\newblock \bibinfo{journal}{\emph{Topics in stroke rehabilitation}}
  \bibinfo{volume}{16}, \bibinfo{number}{2} (\bibinfo{year}{2009}),
  \bibinfo{pages}{133--139}.
\newblock


\bibitem[\protect\citeauthoryear{Michie, Johnston, Abraham, Lawton, Parker, and
  Walker}{Michie et~al\mbox{.}}{2005}]%
        {michie2005making}
\bibfield{author}{\bibinfo{person}{S1 Michie}, \bibinfo{person}{Marie
  Johnston}, \bibinfo{person}{Charles Abraham}, \bibinfo{person}{R Lawton},
  \bibinfo{person}{D Parker}, {and} \bibinfo{person}{A Walker}.}
  \bibinfo{year}{2005}\natexlab{}.
\newblock \showarticletitle{Making psychological theory useful for implementing
  evidence based practice: a consensus approach}.
\newblock \bibinfo{journal}{\emph{BMJ Quality \& Safety}} \bibinfo{volume}{14},
  \bibinfo{number}{1} (\bibinfo{year}{2005}), \bibinfo{pages}{26--33}.
\newblock


\bibitem[\protect\citeauthoryear{Michie, Richardson, Johnston, Abraham,
  Francis, Hardeman, Eccles, Cane, and Wood}{Michie et~al\mbox{.}}{2013}]%
        {michie2013behavior}
\bibfield{author}{\bibinfo{person}{Susan Michie}, \bibinfo{person}{Michelle
  Richardson}, \bibinfo{person}{Marie Johnston}, \bibinfo{person}{Charles
  Abraham}, \bibinfo{person}{Jill Francis}, \bibinfo{person}{Wendy Hardeman},
  \bibinfo{person}{Martin~P Eccles}, \bibinfo{person}{James Cane}, {and}
  \bibinfo{person}{Caroline~E Wood}.} \bibinfo{year}{2013}\natexlab{}.
\newblock \showarticletitle{The behavior change technique taxonomy (v1) of 93
  hierarchically clustered techniques: building an international consensus for
  the reporting of behavior change interventions}.
\newblock \bibinfo{journal}{\emph{Annals of behavioral medicine}}
  \bibinfo{volume}{46}, \bibinfo{number}{1} (\bibinfo{year}{2013}),
  \bibinfo{pages}{81--95}.
\newblock


\bibitem[\protect\citeauthoryear{Middleton, Anton, and Perri}{Middleton
  et~al\mbox{.}}{2013}]%
        {middleton2013long}
\bibfield{author}{\bibinfo{person}{Kathryn~R Middleton},
  \bibinfo{person}{Stephen~D Anton}, {and} \bibinfo{person}{Michal~G Perri}.}
  \bibinfo{year}{2013}\natexlab{}.
\newblock \showarticletitle{Long-term adherence to health behavior change}.
\newblock \bibinfo{journal}{\emph{American journal of lifestyle medicine}}
  \bibinfo{volume}{7}, \bibinfo{number}{6} (\bibinfo{year}{2013}),
  \bibinfo{pages}{395--404}.
\newblock


\bibitem[\protect\citeauthoryear{Moraveji, Akasaka, Pea, and Fogg}{Moraveji
  et~al\mbox{.}}{2011}]%
        {moraveji2011role}
\bibfield{author}{\bibinfo{person}{Neema Moraveji}, \bibinfo{person}{Ryo
  Akasaka}, \bibinfo{person}{Roy Pea}, {and} \bibinfo{person}{BJ Fogg}.}
  \bibinfo{year}{2011}\natexlab{}.
\newblock \showarticletitle{The role of commitment devices and self-shaping in
  persuasive technology}.
\newblock In \bibinfo{booktitle}{\emph{CHI'11 Extended Abstracts on Human
  Factors in Computing Systems}}. \bibinfo{pages}{1591--1596}.
\newblock


\bibitem[\protect\citeauthoryear{Nelder and Wedderburn}{Nelder and
  Wedderburn}{1972}]%
        {nelder1972generalized}
\bibfield{author}{\bibinfo{person}{John~Ashworth Nelder} {and}
  \bibinfo{person}{Robert~WM Wedderburn}.} \bibinfo{year}{1972}\natexlab{}.
\newblock \showarticletitle{Generalized linear models}.
\newblock \bibinfo{journal}{\emph{Journal of the Royal Statistical Society:
  Series A (General)}} \bibinfo{volume}{135}, \bibinfo{number}{3}
  (\bibinfo{year}{1972}), \bibinfo{pages}{370--384}.
\newblock


\bibitem[\protect\citeauthoryear{Oduor and Oinas-Kukkonen}{Oduor and
  Oinas-Kukkonen}{2017}]%
        {oduor2017commitment}
\bibfield{author}{\bibinfo{person}{Michael Oduor} {and} \bibinfo{person}{Harri
  Oinas-Kukkonen}.} \bibinfo{year}{2017}\natexlab{}.
\newblock \showarticletitle{Commitment devices as behavior change support
  systems: a study of users’ perceived competence and continuance intention}.
  In \bibinfo{booktitle}{\emph{International Conference on Persuasive
  Technology}}. Springer, \bibinfo{pages}{201--213}.
\newblock


\bibitem[\protect\citeauthoryear{Okeke, Sobolev, Dell, and Estrin}{Okeke
  et~al\mbox{.}}{2018}]%
        {okeke2018good}
\bibfield{author}{\bibinfo{person}{Fabian Okeke}, \bibinfo{person}{Michael
  Sobolev}, \bibinfo{person}{Nicola Dell}, {and} \bibinfo{person}{Deborah
  Estrin}.} \bibinfo{year}{2018}\natexlab{}.
\newblock \showarticletitle{Good vibrations: can a digital nudge reduce digital
  overload?}. In \bibinfo{booktitle}{\emph{Proceedings of the 20th
  international conference on human-computer interaction with mobile devices
  and services}}. \bibinfo{pages}{1--12}.
\newblock


\bibitem[\protect\citeauthoryear{Pabilonia and Vernon}{Pabilonia and
  Vernon}{2020}]%
        {pabilonia2020telework}
\bibfield{author}{\bibinfo{person}{Sabrina~Wulff Pabilonia} {and}
  \bibinfo{person}{Victoria Vernon}.} \bibinfo{year}{2020}\natexlab{}.
\newblock \showarticletitle{Telework and Time Use in the United States}.
\newblock  (\bibinfo{year}{2020}).
\newblock


\bibitem[\protect\citeauthoryear{Panovich}{Panovich}{2013}]%
        {Panovich13habitbot}
\bibfield{author}{\bibinfo{person}{Katrina Panovich}.}
  \bibinfo{year}{2013}\natexlab{}.
\newblock \bibinfo{title}{Habitbot: A Twitter-based Tool to Help Form and
  Maintain Habits and Goals}.
\newblock
\newblock


\bibitem[\protect\citeauthoryear{Paredes, Gilad-Bachrach, Czerwinski, Roseway,
  Rowan, and Hernandez}{Paredes et~al\mbox{.}}{2014}]%
        {paredes2014poptherapy}
\bibfield{author}{\bibinfo{person}{Pablo Paredes}, \bibinfo{person}{Ran
  Gilad-Bachrach}, \bibinfo{person}{Mary Czerwinski}, \bibinfo{person}{Asta
  Roseway}, \bibinfo{person}{Kael Rowan}, {and} \bibinfo{person}{Javier
  Hernandez}.} \bibinfo{year}{2014}\natexlab{}.
\newblock \showarticletitle{PopTherapy: Coping with stress through
  pop-culture}. In \bibinfo{booktitle}{\emph{Proceedings of the 8th
  International Conference on Pervasive Computing Technologies for
  Healthcare}}. ICST (Institute for Computer Sciences, Social-Informatics
  and~…, \bibinfo{pages}{109--117}.
\newblock


\bibitem[\protect\citeauthoryear{Payne, Sagara, Shu, Appelt, and Johnson}{Payne
  et~al\mbox{.}}{2013}]%
        {payne2013life}
\bibfield{author}{\bibinfo{person}{John~W Payne}, \bibinfo{person}{Namika
  Sagara}, \bibinfo{person}{Suzanne~B Shu}, \bibinfo{person}{Kirstin~C Appelt},
  {and} \bibinfo{person}{Eric~J Johnson}.} \bibinfo{year}{2013}\natexlab{}.
\newblock \showarticletitle{Life expectancy as a constructed belief: Evidence
  of a live-to or die-by framing effect}.
\newblock \bibinfo{journal}{\emph{Journal of Risk and Uncertainty}}
  \bibinfo{volume}{46}, \bibinfo{number}{1} (\bibinfo{year}{2013}),
  \bibinfo{pages}{27--50}.
\newblock


\bibitem[\protect\citeauthoryear{Pe{\~n}a-L{\'o}pez
  et~al\mbox{.}}{Pe{\~n}a-L{\'o}pez et~al\mbox{.}}{2002}]%
        {pena2002nation}
\bibfield{author}{\bibinfo{person}{Ismael Pe{\~n}a-L{\'o}pez} {et~al\mbox{.}}}
  \bibinfo{year}{2002}\natexlab{}.
\newblock \showarticletitle{A nation online: How Americans are expanding their
  use of the Internet}.
\newblock  (\bibinfo{year}{2002}).
\newblock


\bibitem[\protect\citeauthoryear{Peters, Dieckmann, V{\"a}stfj{\"a}ll, Mertz,
  Slovic, and Hibbard}{Peters et~al\mbox{.}}{2009}]%
        {peters2009bringing}
\bibfield{author}{\bibinfo{person}{Ellen Peters}, \bibinfo{person}{Nathan~F
  Dieckmann}, \bibinfo{person}{Daniel V{\"a}stfj{\"a}ll}, \bibinfo{person}{CK
  Mertz}, \bibinfo{person}{Paul Slovic}, {and} \bibinfo{person}{Judith~H
  Hibbard}.} \bibinfo{year}{2009}\natexlab{}.
\newblock \showarticletitle{Bringing meaning to numbers: The impact of
  evaluative categories on decisions.}
\newblock \bibinfo{journal}{\emph{Journal of experimental psychology: applied}}
  \bibinfo{volume}{15}, \bibinfo{number}{3} (\bibinfo{year}{2009}),
  \bibinfo{pages}{213}.
\newblock


\bibitem[\protect\citeauthoryear{Prochaska and Velicer}{Prochaska and
  Velicer}{1997}]%
        {prochaska1997transtheoretical}
\bibfield{author}{\bibinfo{person}{James~O Prochaska} {and}
  \bibinfo{person}{Wayne~F Velicer}.} \bibinfo{year}{1997}\natexlab{}.
\newblock \showarticletitle{The transtheoretical model of health behavior
  change}.
\newblock \bibinfo{journal}{\emph{American journal of health promotion}}
  \bibinfo{volume}{12}, \bibinfo{number}{1} (\bibinfo{year}{1997}),
  \bibinfo{pages}{38--48}.
\newblock


\bibitem[\protect\citeauthoryear{Quoquab, Salam, and Halimah}{Quoquab
  et~al\mbox{.}}{2015}]%
        {quoquab2015does}
\bibfield{author}{\bibinfo{person}{Farzana Quoquab},
  \bibinfo{person}{Zarina~Abdul Salam}, {and} \bibinfo{person}{Siti Halimah}.}
  \bibinfo{year}{2015}\natexlab{}.
\newblock \showarticletitle{Does cyberloafing boost employee productivity?}. In
  \bibinfo{booktitle}{\emph{2015 International Symposium on Technology
  Management and Emerging Technologies (ISTMET)}}. IEEE,
  \bibinfo{pages}{119--122}.
\newblock


\bibitem[\protect\citeauthoryear{Racey, O'Brien, Douglas, Marquez, Hendrie, and
  Newton}{Racey et~al\mbox{.}}{2016}]%
        {racey2016systematic}
\bibfield{author}{\bibinfo{person}{Megan Racey}, \bibinfo{person}{Charlene
  O'Brien}, \bibinfo{person}{Sabrina Douglas}, \bibinfo{person}{Olivia
  Marquez}, \bibinfo{person}{Gilly Hendrie}, {and} \bibinfo{person}{Genevieve
  Newton}.} \bibinfo{year}{2016}\natexlab{}.
\newblock \showarticletitle{Systematic review of school-based interventions to
  modify dietary behavior: does intervention intensity impact effectiveness?}
\newblock \bibinfo{journal}{\emph{Journal of School Health}}
  \bibinfo{volume}{86}, \bibinfo{number}{6} (\bibinfo{year}{2016}),
  \bibinfo{pages}{452--463}.
\newblock


\bibitem[\protect\citeauthoryear{Reinecke, Vorderer, and Knop}{Reinecke
  et~al\mbox{.}}{2014}]%
        {reinecke2014entertainment}
\bibfield{author}{\bibinfo{person}{Leonard Reinecke}, \bibinfo{person}{Peter
  Vorderer}, {and} \bibinfo{person}{Katharina Knop}.}
  \bibinfo{year}{2014}\natexlab{}.
\newblock \showarticletitle{Entertainment 2.0? The role of intrinsic and
  extrinsic need satisfaction for the enjoyment of Facebook use}.
\newblock \bibinfo{journal}{\emph{Journal of Communication}}
  \bibinfo{volume}{64}, \bibinfo{number}{3} (\bibinfo{year}{2014}),
  \bibinfo{pages}{417--438}.
\newblock


\bibitem[\protect\citeauthoryear{Rhodes and Courneya}{Rhodes and
  Courneya}{2004}]%
        {rhodes2004differentiating}
\bibfield{author}{\bibinfo{person}{RE Rhodes} {and} \bibinfo{person}{KS
  Courneya}.} \bibinfo{year}{2004}\natexlab{}.
\newblock \showarticletitle{Differentiating motivation and control in the
  theory of planned behavior}.
\newblock \bibinfo{journal}{\emph{Psychology, Health \& Medicine}}
  \bibinfo{volume}{9}, \bibinfo{number}{2} (\bibinfo{year}{2004}),
  \bibinfo{pages}{205--215}.
\newblock


\bibitem[\protect\citeauthoryear{Rosen}{Rosen}{2000}]%
        {rosen2000sequencing}
\bibfield{author}{\bibinfo{person}{Craig~S Rosen}.}
  \bibinfo{year}{2000}\natexlab{}.
\newblock \showarticletitle{Is the sequencing of change processes by stage
  consistent across health problems? A meta-analysis.}
\newblock \bibinfo{journal}{\emph{Health psychology}} \bibinfo{volume}{19},
  \bibinfo{number}{6} (\bibinfo{year}{2000}), \bibinfo{pages}{593}.
\newblock


\bibitem[\protect\citeauthoryear{Rosenzweig, Harackiewicz, Priniski, Hecht,
  Canning, Tibbetts, and Hyde}{Rosenzweig et~al\mbox{.}}{2019}]%
        {rosenzweig2019choose}
\bibfield{author}{\bibinfo{person}{Emily~Q Rosenzweig},
  \bibinfo{person}{Judith~M Harackiewicz}, \bibinfo{person}{Stacy~J Priniski},
  \bibinfo{person}{Cameron~A Hecht}, \bibinfo{person}{Elizabeth~A Canning},
  \bibinfo{person}{Yoi Tibbetts}, {and} \bibinfo{person}{Janet~S Hyde}.}
  \bibinfo{year}{2019}\natexlab{}.
\newblock \showarticletitle{Choose your own intervention: Using choice to
  enhance the effectiveness of a utility-value intervention.}
\newblock \bibinfo{journal}{\emph{Motivation Science}} \bibinfo{volume}{5},
  \bibinfo{number}{3} (\bibinfo{year}{2019}), \bibinfo{pages}{269}.
\newblock


\bibitem[\protect\citeauthoryear{Ryan, Patrick, Deci, and Williams}{Ryan
  et~al\mbox{.}}{2008}]%
        {ryan2008facilitating}
\bibfield{author}{\bibinfo{person}{Richard~M Ryan}, \bibinfo{person}{Heather
  Patrick}, \bibinfo{person}{Edward~L Deci}, {and} \bibinfo{person}{Geoffrey~C
  Williams}.} \bibinfo{year}{2008}\natexlab{}.
\newblock \showarticletitle{Facilitating health behaviour change and its
  maintenance: Interventions based on self-determination theory}.
\newblock \bibinfo{journal}{\emph{The European health psychologist}}
  \bibinfo{volume}{10}, \bibinfo{number}{1} (\bibinfo{year}{2008}),
  \bibinfo{pages}{2--5}.
\newblock


\bibitem[\protect\citeauthoryear{Samuelson and Zeckhauser}{Samuelson and
  Zeckhauser}{1988}]%
        {samuelson1988status}
\bibfield{author}{\bibinfo{person}{William Samuelson} {and}
  \bibinfo{person}{Richard Zeckhauser}.} \bibinfo{year}{1988}\natexlab{}.
\newblock \showarticletitle{Status quo bias in decision making}.
\newblock \bibinfo{journal}{\emph{Journal of risk and uncertainty}}
  \bibinfo{volume}{1}, \bibinfo{number}{1} (\bibinfo{year}{1988}),
  \bibinfo{pages}{7--59}.
\newblock


\bibitem[\protect\citeauthoryear{Scharkow}{Scharkow}{2016}]%
        {scharkow2016accuracy}
\bibfield{author}{\bibinfo{person}{Michael Scharkow}.}
  \bibinfo{year}{2016}\natexlab{}.
\newblock \showarticletitle{The accuracy of self-reported internet use—A
  validation study using client log data}.
\newblock \bibinfo{journal}{\emph{Communication Methods and Measures}}
  \bibinfo{volume}{10}, \bibinfo{number}{1} (\bibinfo{year}{2016}),
  \bibinfo{pages}{13--27}.
\newblock


\bibitem[\protect\citeauthoryear{Schueller}{Schueller}{2010}]%
        {schueller2010preferences}
\bibfield{author}{\bibinfo{person}{Stephen~M Schueller}.}
  \bibinfo{year}{2010}\natexlab{}.
\newblock \showarticletitle{Preferences for positive psychology exercises}.
\newblock \bibinfo{journal}{\emph{The Journal of Positive Psychology}}
  \bibinfo{volume}{5}, \bibinfo{number}{3} (\bibinfo{year}{2010}),
  \bibinfo{pages}{192--203}.
\newblock


\bibitem[\protect\citeauthoryear{Schwarzer}{Schwarzer}{2008}]%
        {schwarzer2008modeling}
\bibfield{author}{\bibinfo{person}{Ralf Schwarzer}.}
  \bibinfo{year}{2008}\natexlab{}.
\newblock \showarticletitle{Modeling health behavior change: How to predict and
  modify the adoption and maintenance of health behaviors}.
\newblock \bibinfo{journal}{\emph{Applied psychology}} \bibinfo{volume}{57},
  \bibinfo{number}{1} (\bibinfo{year}{2008}), \bibinfo{pages}{1--29}.
\newblock


\bibitem[\protect\citeauthoryear{Scollon, Prieto, and Diener}{Scollon
  et~al\mbox{.}}{2009}]%
        {scollon2009experience}
\bibfield{author}{\bibinfo{person}{Christie~Napa Scollon},
  \bibinfo{person}{Chu-Kim Prieto}, {and} \bibinfo{person}{Ed Diener}.}
  \bibinfo{year}{2009}\natexlab{}.
\newblock \showarticletitle{Experience sampling: promises and pitfalls,
  strength and weaknesses}.
\newblock In \bibinfo{booktitle}{\emph{Assessing well-being}}.
  \bibinfo{publisher}{Springer}, \bibinfo{pages}{157--180}.
\newblock


\bibitem[\protect\citeauthoryear{Sharpe}{Sharpe}{2015}]%
        {sharpe2015chi}
\bibfield{author}{\bibinfo{person}{Donald Sharpe}.}
  \bibinfo{year}{2015}\natexlab{}.
\newblock \showarticletitle{Chi-Square Test is Statistically Significant: Now
  What?}
\newblock \bibinfo{journal}{\emph{Practical Assessment, Research, and
  Evaluation}} \bibinfo{volume}{20}, \bibinfo{number}{1}
  (\bibinfo{year}{2015}), \bibinfo{pages}{8}.
\newblock


\bibitem[\protect\citeauthoryear{Shrout and Fleiss}{Shrout and Fleiss}{1979}]%
        {shrout1979intraclass}
\bibfield{author}{\bibinfo{person}{Patrick~E Shrout} {and}
  \bibinfo{person}{Joseph~L Fleiss}.} \bibinfo{year}{1979}\natexlab{}.
\newblock \showarticletitle{Intraclass correlations: uses in assessing rater
  reliability.}
\newblock \bibinfo{journal}{\emph{Psychological bulletin}}
  \bibinfo{volume}{86}, \bibinfo{number}{2} (\bibinfo{year}{1979}),
  \bibinfo{pages}{420}.
\newblock


\bibitem[\protect\citeauthoryear{Shu}{Shu}{2008}]%
        {shu2008future}
\bibfield{author}{\bibinfo{person}{Suzanne~B Shu}.}
  \bibinfo{year}{2008}\natexlab{}.
\newblock \showarticletitle{Future-biased search: the quest for the ideal}.
\newblock \bibinfo{journal}{\emph{Journal of Behavioral Decision Making}}
  \bibinfo{volume}{21}, \bibinfo{number}{4} (\bibinfo{year}{2008}),
  \bibinfo{pages}{352--377}.
\newblock


\bibitem[\protect\citeauthoryear{Shu and Gneezy}{Shu and Gneezy}{2010}]%
        {shu2010procrastination}
\bibfield{author}{\bibinfo{person}{Suzanne~B Shu} {and} \bibinfo{person}{Ayelet
  Gneezy}.} \bibinfo{year}{2010}\natexlab{}.
\newblock \showarticletitle{Procrastination of enjoyable experiences}.
\newblock \bibinfo{journal}{\emph{Journal of Marketing Research}}
  \bibinfo{volume}{47}, \bibinfo{number}{5} (\bibinfo{year}{2010}),
  \bibinfo{pages}{933--944}.
\newblock


\bibitem[\protect\citeauthoryear{Simonson}{Simonson}{1990}]%
        {simonson1990effect}
\bibfield{author}{\bibinfo{person}{Itamar Simonson}.}
  \bibinfo{year}{1990}\natexlab{}.
\newblock \showarticletitle{The effect of purchase quantity and timing on
  variety-seeking behavior}.
\newblock \bibinfo{journal}{\emph{Journal of Marketing Research}}
  \bibinfo{volume}{27}, \bibinfo{number}{2} (\bibinfo{year}{1990}),
  \bibinfo{pages}{150--162}.
\newblock


\bibitem[\protect\citeauthoryear{Soll, Keeney, and Larrick}{Soll
  et~al\mbox{.}}{2013}]%
        {soll2013consumer}
\bibfield{author}{\bibinfo{person}{Jack~B Soll}, \bibinfo{person}{Ralph~L
  Keeney}, {and} \bibinfo{person}{Richard~P Larrick}.}
  \bibinfo{year}{2013}\natexlab{}.
\newblock \showarticletitle{Consumer misunderstanding of credit card use,
  payments, and debt: Causes and solutions}.
\newblock \bibinfo{journal}{\emph{Journal of Public Policy \& Marketing}}
  \bibinfo{volume}{32}, \bibinfo{number}{1} (\bibinfo{year}{2013}),
  \bibinfo{pages}{66--81}.
\newblock


\bibitem[\protect\citeauthoryear{Soman, Ainslie, Frederick, Li, Lynch, Moreau,
  Mitchell, Read, Sawyer, Trope, et~al\mbox{.}}{Soman et~al\mbox{.}}{2005}]%
        {soman2005psychology}
\bibfield{author}{\bibinfo{person}{Dilip Soman}, \bibinfo{person}{George
  Ainslie}, \bibinfo{person}{Shane Frederick}, \bibinfo{person}{Xiuping Li},
  \bibinfo{person}{John Lynch}, \bibinfo{person}{Page Moreau},
  \bibinfo{person}{Andrew Mitchell}, \bibinfo{person}{Daniel Read},
  \bibinfo{person}{Alan Sawyer}, \bibinfo{person}{Yaacov Trope},
  {et~al\mbox{.}}} \bibinfo{year}{2005}\natexlab{}.
\newblock \showarticletitle{The psychology of intertemporal discounting: Why
  are distant events valued differently from proximal ones?}
\newblock \bibinfo{journal}{\emph{Marketing Letters}} \bibinfo{volume}{16},
  \bibinfo{number}{3-4} (\bibinfo{year}{2005}), \bibinfo{pages}{347--360}.
\newblock


\bibitem[\protect\citeauthoryear{Sparks, Shepherd, Wieringa, and
  Zimmermanns}{Sparks et~al\mbox{.}}{1995}]%
        {sparks1995perceived}
\bibfield{author}{\bibinfo{person}{Paul Sparks}, \bibinfo{person}{Richard
  Shepherd}, \bibinfo{person}{Nicole Wieringa}, {and} \bibinfo{person}{Nicole
  Zimmermanns}.} \bibinfo{year}{1995}\natexlab{}.
\newblock \showarticletitle{Perceived behavioural control, unrealistic optimism
  and dietary change: An exploratory study}.
\newblock \bibinfo{journal}{\emph{Appetite}} \bibinfo{volume}{24},
  \bibinfo{number}{3} (\bibinfo{year}{1995}), \bibinfo{pages}{243--255}.
\newblock


\bibitem[\protect\citeauthoryear{Teixeira, Marques, Silva, Brunet, Duda,
  Haerens, La~Guardia, Lindwall, Lonsdale, Markland, et~al\mbox{.}}{Teixeira
  et~al\mbox{.}}{2020}]%
        {teixeira2020classification}
\bibfield{author}{\bibinfo{person}{Pedro~J Teixeira}, \bibinfo{person}{Marta~M
  Marques}, \bibinfo{person}{Marlene~N Silva}, \bibinfo{person}{Jennifer
  Brunet}, \bibinfo{person}{Joan~L Duda}, \bibinfo{person}{Leen Haerens},
  \bibinfo{person}{Jennifer La~Guardia}, \bibinfo{person}{Magnus Lindwall},
  \bibinfo{person}{Chris Lonsdale}, \bibinfo{person}{David Markland},
  {et~al\mbox{.}}} \bibinfo{year}{2020}\natexlab{}.
\newblock \showarticletitle{A classification of motivation and behavior change
  techniques used in self-determination theory-based interventions in health
  contexts.}
\newblock \bibinfo{journal}{\emph{Motivation Science}} \bibinfo{volume}{6},
  \bibinfo{number}{4} (\bibinfo{year}{2020}), \bibinfo{pages}{438}.
\newblock


\bibitem[\protect\citeauthoryear{Thaler}{Thaler}{1980}]%
        {thaler1980toward}
\bibfield{author}{\bibinfo{person}{Richard Thaler}.}
  \bibinfo{year}{1980}\natexlab{}.
\newblock \showarticletitle{Toward a positive theory of consumer choice}.
\newblock \bibinfo{journal}{\emph{Journal of Economic Behavior \&
  Organization}} \bibinfo{volume}{1}, \bibinfo{number}{1}
  (\bibinfo{year}{1980}), \bibinfo{pages}{39--60}.
\newblock


\bibitem[\protect\citeauthoryear{Toma and Hancock}{Toma and Hancock}{2013}]%
        {toma2013self}
\bibfield{author}{\bibinfo{person}{Catalina~L Toma} {and}
  \bibinfo{person}{Jeffrey~T Hancock}.} \bibinfo{year}{2013}\natexlab{}.
\newblock \showarticletitle{Self-affirmation underlies Facebook use}.
\newblock \bibinfo{journal}{\emph{Personality and Social Psychology Bulletin}}
  \bibinfo{volume}{39}, \bibinfo{number}{3} (\bibinfo{year}{2013}),
  \bibinfo{pages}{321--331}.
\newblock


\bibitem[\protect\citeauthoryear{Tromholt}{Tromholt}{2016}]%
        {tromholt2016facebook}
\bibfield{author}{\bibinfo{person}{Morten Tromholt}.}
  \bibinfo{year}{2016}\natexlab{}.
\newblock \showarticletitle{The Facebook experiment: Quitting Facebook leads to
  higher levels of well-being}.
\newblock \bibinfo{journal}{\emph{Cyberpsychology, behavior, and social
  networking}} \bibinfo{volume}{19}, \bibinfo{number}{11}
  (\bibinfo{year}{2016}), \bibinfo{pages}{661--666}.
\newblock


\bibitem[\protect\citeauthoryear{Tseng, Lee, Denoue, and Avrahami}{Tseng
  et~al\mbox{.}}{2019}]%
        {tseng2019overcoming}
\bibfield{author}{\bibinfo{person}{Vincent W-S Tseng},
  \bibinfo{person}{Matthew~L Lee}, \bibinfo{person}{Laurent Denoue}, {and}
  \bibinfo{person}{Daniel Avrahami}.} \bibinfo{year}{2019}\natexlab{}.
\newblock \showarticletitle{Overcoming Distractions during Transitions from
  Break to Work using a Conversational Website-Blocking System}. In
  \bibinfo{booktitle}{\emph{Proceedings of the 2019 CHI Conference on Human
  Factors in Computing Systems}}. \bibinfo{pages}{1--13}.
\newblock


\bibitem[\protect\citeauthoryear{Van~Cappellen, Catalino, and
  Fredrickson}{Van~Cappellen et~al\mbox{.}}{2019}]%
        {van2019new}
\bibfield{author}{\bibinfo{person}{Patty Van~Cappellen},
  \bibinfo{person}{Lahnna~I Catalino}, {and} \bibinfo{person}{Barbara~L
  Fredrickson}.} \bibinfo{year}{2019}\natexlab{}.
\newblock \showarticletitle{A new micro-intervention to increase the enjoyment
  and continued practice of meditation.}
\newblock \bibinfo{journal}{\emph{Emotion}} (\bibinfo{year}{2019}).
\newblock


\bibitem[\protect\citeauthoryear{Vitak, Crouse, and LaRose}{Vitak
  et~al\mbox{.}}{2011}]%
        {vitak2011personal}
\bibfield{author}{\bibinfo{person}{Jessica Vitak}, \bibinfo{person}{Julia
  Crouse}, {and} \bibinfo{person}{Robert LaRose}.}
  \bibinfo{year}{2011}\natexlab{}.
\newblock \showarticletitle{Personal Internet use at work: Understanding
  cyberslacking}.
\newblock \bibinfo{journal}{\emph{Computers in Human Behavior}}
  \bibinfo{volume}{27}, \bibinfo{number}{5} (\bibinfo{year}{2011}),
  \bibinfo{pages}{1751--1759}.
\newblock


\bibitem[\protect\citeauthoryear{Vohs, Baumeister, Schmeichel, Twenge, Nelson,
  and Tice}{Vohs et~al\mbox{.}}{2014}]%
        {vohs2014making}
\bibfield{author}{\bibinfo{person}{Kathleen~D Vohs}, \bibinfo{person}{Roy~F
  Baumeister}, \bibinfo{person}{Brandon~J Schmeichel}, \bibinfo{person}{Jean~M
  Twenge}, \bibinfo{person}{Noelle~M Nelson}, {and} \bibinfo{person}{Dianne~M
  Tice}.} \bibinfo{year}{2014}\natexlab{}.
\newblock \showarticletitle{Making choices impairs subsequent self-control: a
  limited-resource account of decision making, self-regulation, and active
  initiative.}
\newblock  (\bibinfo{year}{2014}).
\newblock


\bibitem[\protect\citeauthoryear{Weber, Johnson, Milch, Chang, Brodscholl, and
  Goldstein}{Weber et~al\mbox{.}}{2007}]%
        {weber2007asymmetric}
\bibfield{author}{\bibinfo{person}{Elke~U Weber}, \bibinfo{person}{Eric~J
  Johnson}, \bibinfo{person}{Kerry~F Milch}, \bibinfo{person}{Hannah Chang},
  \bibinfo{person}{Jeffrey~C Brodscholl}, {and} \bibinfo{person}{Daniel~G
  Goldstein}.} \bibinfo{year}{2007}\natexlab{}.
\newblock \showarticletitle{Asymmetric discounting in intertemporal choice: A
  query-theory account}.
\newblock \bibinfo{journal}{\emph{Psychological science}} \bibinfo{volume}{18},
  \bibinfo{number}{6} (\bibinfo{year}{2007}), \bibinfo{pages}{516--523}.
\newblock


\bibitem[\protect\citeauthoryear{Yanay and Yanay}{Yanay and Yanay}{2008}]%
        {yanay2008decline}
\bibfield{author}{\bibinfo{person}{Galit~Ventura Yanay} {and}
  \bibinfo{person}{Niza Yanay}.} \bibinfo{year}{2008}\natexlab{}.
\newblock \showarticletitle{The decline of motivation?: From commitment to
  dropping out of volunteering}.
\newblock \bibinfo{journal}{\emph{Nonprofit management and Leadership}}
  \bibinfo{volume}{19}, \bibinfo{number}{1} (\bibinfo{year}{2008}),
  \bibinfo{pages}{65--78}.
\newblock


\bibitem[\protect\citeauthoryear{Yeung}{Yeung}{2004}]%
        {yeung2004octagon}
\bibfield{author}{\bibinfo{person}{Anne~Birgitta Yeung}.}
  \bibinfo{year}{2004}\natexlab{}.
\newblock \showarticletitle{The octagon model of volunteer motivation: Results
  of a phenomenological analysis}.
\newblock \bibinfo{journal}{\emph{Voluntas: International Journal of Voluntary
  and Nonprofit Organizations}} \bibinfo{volume}{15}, \bibinfo{number}{1}
  (\bibinfo{year}{2004}), \bibinfo{pages}{21--46}.
\newblock


\bibitem[\protect\citeauthoryear{Zauberman and Lynch~Jr}{Zauberman and
  Lynch~Jr}{2005}]%
        {zauberman2005resource}
\bibfield{author}{\bibinfo{person}{Gal Zauberman} {and} \bibinfo{person}{John~G
  Lynch~Jr}.} \bibinfo{year}{2005}\natexlab{}.
\newblock \showarticletitle{Resource slack and propensity to discount delayed
  investments of time versus money.}
\newblock \bibinfo{journal}{\emph{Journal of Experimental Psychology: General}}
  \bibinfo{volume}{134}, \bibinfo{number}{1} (\bibinfo{year}{2005}),
  \bibinfo{pages}{23}.
\newblock


\bibitem[\protect\citeauthoryear{Zhao, Liu, Wu, Yao, and Huang}{Zhao
  et~al\mbox{.}}{2018}]%
        {zhao2018watch}
\bibfield{author}{\bibinfo{person}{Huaiyi Zhao}, \bibinfo{person}{Jie Liu},
  \bibinfo{person}{Jiahao Wu}, \bibinfo{person}{Kaiyu Yao}, {and}
  \bibinfo{person}{Jin Huang}.} \bibinfo{year}{2018}\natexlab{}.
\newblock \showarticletitle{Watch-Learning: Using the Smartwatch for Secondary
  Language Vocabulary Learning}. In \bibinfo{booktitle}{\emph{Proceedings of
  the Sixth International Symposium of Chinese CHI}}.
  \bibinfo{pages}{108--111}.
\newblock


\end{thebibliography}
